%% file: manuscript.tex
\ifx\onecol\undefined
\documentclass[journal]{IEEEtran}
\else 
\documentclass[onecolumn,draftcls,12pt]{IEEEtran}
\fi

\usepackage[utf8]{inputenc}
\usepackage{physics}
\usepackage{amsmath}
\usepackage{amssymb}
\usepackage{subfigure}
\usepackage{graphicx}
\usepackage{graphics}
\usepackage{color}
\usepackage{psfrag}
\usepackage{cite}
\usepackage{balance}
\usepackage{algorithm}
\usepackage{accents}
\usepackage{amsthm}
\usepackage{bm}
\usepackage{url}
\usepackage{algorithmic}
\usepackage[english]{babel}
\usepackage{multirow}
\usepackage{enumerate}
\usepackage{cases}
\usepackage{stfloats}
\usepackage{dsfont}
\usepackage{soul}
\usepackage{amsfonts}
\usepackage{fancyhdr}
\usepackage{hhline}
\usepackage{array}
\usepackage{booktabs}
\usepackage{framed}
\usepackage{float}
\usepackage{soul}
\usepackage{url}
\usepackage{hyperref}
\usepackage{mathtools}

\hypersetup{hidelinks}

\newcommand{\mb}[1]{{  \mathbf  #1}}

\begin{document}
\include{header}

\title{A Theory of Atomic Beamforming
}
\author{{Mingyao Cui, Qunsong Zeng, and Kaibin Huang,~\IEEEmembership{Fellow, IEEE}}
\thanks{M. Cui, Q. Zeng, and K. Huang are with the Department of Electrical and Computer Engineering, The University of Hong Kong, Hong Kong (Emails: \{mycui, qszeng\}@eee.hku.hk, huangkb@hku.hk). Corresponding authors: Q. Zeng; K. Huang.}
}
\maketitle

\begin{abstract}
Leveraging the quantum advantages of highly excited atoms, Rydberg atomic receivers (RAREs) represent a paradigm shift in microwave detection with extremely high sensitivity and broadband tunability. 
However, existing studies often model RAREs as isotropic point receivers and neglect the spatial variation of Rydberg atomic states within vapor cells, which can lead to inaccurate characterization of their reception patterns. 
To address this issue, we theoretically analyze the spatial response of a local-oscillator (LO) field-dressed RARE. 
Our results reveal that as the vapor-cell length increases, a receiving beam aligned with the LO field is formed, and its beamwidth is inversely proportional to the cell length.
This finding enables atomic beamforming with only a single atomic vapor cell to enhance signal-to-noise ratio.  Furthermore, we analyze the maximum beamforming gain of a single vapor cell by balancing the fundamental tradeoff between improved spatial selectivity and increased laser attenuation in the atomic medium as the vapor cell length becomes longer.
To mitigate the beamforming-gain loss caused by laser attenuation, we further propose a segmental-vapor-cell architecture. In this architecture, multiple short vapor cells are arranged along the optical propagation path and separated by clear-air gaps. This design effectively expands the reception aperture while keeping the total vapor-cell length,  and hence attenuation loss, fixed. 
As a result, it achieves a narrower beamwidth and higher beamforming gain than a conventional single-vapor-cell receiver, as demonstrated by extensive numerical results.

\end{abstract}


\section{Introduction}
The rapid advancement in quantum information science and technology is poised to revolutionize the fields of computation, communication, and sensing~\cite{QIST_Hanzo2025}. 
Among these, quantum sensing stands out as particularly promising due to its potential to enable a wide range of applications and thereby significantly impact society~\cite{QuanSense_Degen2017, QuanSense_Zhang2021}. 
By harnessing quantum phenomena such as entanglement and quantum state squeezing, these technologies offer measurement sensitivities that surpass classical limits. 
The paradigm shift has led to the development of various high-precision quantum sensors, including magnetometers, gravimeters, and electrometers, which have propelled advances in physics, metrology, geophysics, and communications~\cite{QuanSense_Degen2017}. 
Building upon these quantum advancements, Rydberg atomic receivers (RAREs) are expected to influence wireless communication by enabling high-precision detection of radio-frequency (RF) electric fields~\cite{ RydbMag_Cui2025}.
The focus of this work is the development of novel RARE-enabled wireless communication techniques.

RAREs operate on quantum-mechanical principles that are fundamentally different from those of classical RF receivers~\cite{fan_atom_2015}. 
Specifically, rather than using metallic antennas to transduce RF signals into electrical currents, RAREs employ \emph{quantum antennas}, referring to alkali-metal atoms with highly excited Rydberg states.
When exposed to external RF fields, these atoms undergo transitions between Rydberg energy levels, inducing measurable shifts in their optical transmission spectra.
To detect these shifts, laser beams are propagated through a vapor cell containing these Rydberg atoms, and the output light is monitored using photodetectors (PDs)~\cite{sedlacek_microwave_2012}.
This all-optical transduction enables direct RF-to-baseband conversion without the need for traditional electronic front-end components. 
As a result, RAREs bypass
the thermal noise limits imposed by Johnson–Nyquist noise~\cite{nyquist_thermal_1928}. 
Theoretically, RAREs can approach the standard quantum limit (SQL) of sensitivity, approximately $700\:{\rm pV\cdot {cm}^{-1}\cdot{Hz}^{-1/2}}$~\cite{RydNP_Jing2020}, which is lower than the thermal noise floor of about $0.98\:{\rm nV \cdot{cm}^{-1}\cdot{Hz}^{-1/2}}$~\cite{Resonant_Sandidge2024}.
In addition to their high sensitivity, RAREs offer several other advantages. On one hand, a single RARE 
device can provide ultra-broadband coverage---from DC up to the terahertz regime---by leveraging the so-called AC Stark shift and the dense atomic energy levels~\cite{LowRydberg_Li2023, Multiband_Cui2025}.
In contrast, classical receivers require multiple band-specific hardware components to detect signals across different bands. 
On the other hand, when configured as an array, RAREs exhibit negligible mutual coupling, which is an issue that commonly affects traditional arrays. This is attributed to the minimal interaction between spatially separated laser beams~\cite{RydbMag_Cui2025}.
With these distinctive features, the deployment of RAREs in next-generation wireless systems promises to achieve unprecedented sensitivity, connectivity, and spectral efficiency.

Though research on RAREs remains in its early stages, recent efforts have significantly advanced the field along three primary directions.
First, there has been continuous progress in the development of RARE architectures, aimed at demonstrating signal detection capability, enhancing sensitivity, and improving system compactness. 
The foundational RARE architecture relies on two quantum {\color{black}phenomena}, electromagnetically induced transparency (EIT) and Autler-Townes (AT) splitting~\cite{sedlacek_microwave_2012, Zushiye_Holloway2010}, achieving sensitivities on the order of $\sim$${\rm \mu V\cdot {cm}^{-1}\cdot{Hz}^{-1/2}}$.
The need for frequency scanning of a laser beam for observing EIT-AT can be eliminated by incorporating an external local oscillator (LO) field, leading to the \emph{superheterodyne} architecture~\cite{RydNP_Jing2020}. 
Such LO-dressed RARE systems have achieved remarkable sensitivity of approximately $1{\rm n V \cdot{cm}^{-1}\cdot{Hz}^{-1/2}}$~\cite{ RydNP_Jing2020, cai_sensitivity_2023}. 
Most recently, the \emph{self-heterodyne} detection architecture has been proposed for high-resolution radar ranging~\cite{QWS_Cui2025}. Its key innovation lies in replacing the external LO with the transmitted signal itself to enhance system compactness.
Second, RARE-based wireless communication techniques have been designed to support a variety of modulation schemes, including amplitude, frequency, and phase modulation~\cite{RydPhase_Meyer2018, RydPhase_Simons2019}. 
Advances in this area have led to the development of sophisticated RARE systems~\cite{AtomicMIMO_Cui2025,RydMIMO_Sha2025,liu_continuous-frequency_2022, Multiband_Cui2025, RydMultiband_Du2022}.
Notably, the concept of RARE arrays has been introduced to enable the detection of spatially multiplexed signals and pave the way for a novel paradigm termed atomic multiple-input-multiple-output (atomic-MIMO) communications~\cite{AtomicMIMO_Cui2025, RydMIMO_Sha2025}. Additionally, RAREs have been designed to support both continuous-band and multi-band reception in broadband communication systems~\cite{liu_continuous-frequency_2022, Multiband_Cui2025, RydMultiband_Du2022}.
The third direction focuses on leveraging RAREs for a wide range of sensing applications. 
These include angle-of-arrival (AoA) estimation~\cite{RydAOA_Robinson2021, RydbergAoA_Guo2025}, displacement detection~\cite{QuanSense_Zhang2023}, multi-band localization~\cite{QuanSense_Chen2025}, moisture sensing~\cite{PhysRevApplied.21.044025}, and integrated sensing and communication~\cite{RydISAC_chen}.


State-of-the-art RARE designs are predominantly based on the assumption that atomic vapor cells exhibit an isotropic response to incident RF fields~\cite{RydNP_Jing2020, Rydberg_Gong2025, liu_continuous-frequency_2022, Multiband_Cui2025, Dynamic_Zhu2025, RydMIMO_Sha2025, IsoRyd_Jia}.
This assumption, which is justified when the vapor cell is sufficiently short, allows the RARE to be modeled as a point receiver, thereby simplifying the input-output signal model that underpins detection and signal processing. 
Specifically, a signal model for LO-dressed RAREs was initially established by authors of~\cite{RydNP_Jing2020} via solving the Lindblad master equation in the steady state. 
Subsequent studies have extended this simplified model to accommodate systems involving off-resonant optical and RF fields~\cite{RydChen_Gong2025, Rydberg_Gong2025, liu_continuous-frequency_2022}, as well as to enable multi-band signal reception~\cite{Multiband_Cui2025}. 
Further advancements have generalized the steady model~\cite{RydNP_Jing2020} to account for dynamic signals through Laplace transform analysis of quantum state evolution~\cite{Dynamic_Zhu2025}. 
As a result of the isotropic approximation, these models often treat RAREs analogously to isotropic antennas. 
However, in practical scenarios, the atomic quantum states within the vapor cell typically exhibit spatial variation, especially in LO-dressed RAREs. 
This spatial inhomogeneity can significantly influence both the response pattern and the antenna gain of the RARE. 
Therefore, characterization of these spatial effects is essential for a complete understanding of RARE operation so as to fully realize their potential. 
However, such in-depth spatial analyses remain missing in the existing literature.
To fill this gap, we theoretically analyze the effects of spatially varying quantum states of Rydberg atoms inside the vapor cell. 
{Our study leads to the discovery of atomic beamforming: a directional (receive) beam can be generated by a long atomic vapor cell with only a single pair of laser beams, a single PD, and a single LO.}
The main contributions are summarized below.

\begin{itemize}
    \item \textbf{Atomic beamforming with a single vapor cell}: 
    Consider the standard RARE with a single atomic vapor cell. 
    We derive a new signal model that accounts for the spatial variation of quantum states induced by the interference between the incident signal and LO fields. 
    A key insight from this analysis is that a vapor cell with substantial length exhibits a directional reception pattern. 
    This pattern naturally steers a beam toward the LO field direction with a beamwidth inversely proportional to the cell length, effectively realizing single-vapor-cell atomic beamforming. 
    To further explore this phenomenon, we analyze the achievable beamforming gain under two noise regimes: the blackbody radiation (BBR)-limited regime and the photon shot noise (PSN)-limited regime. 
    Our results show that, in the BBR regime, the beamforming gain increases linearly with the cell length, whereas in the PSN regime, it decays exponentially due to severe laser attenuation inside the atomic medium. 
   This tradeoff implies the existence of an optimal cell length that maximizes the beamforming gain, for which we derive an analytical expression.

    \item \textbf{Atomic beamforming with segmental vapor cells}:  
    To overcome the exponential decay in beamforming gain with increasing cell length in the PSN regime, we propose a novel segmental-vapor-cell architecture for Rydberg receivers.
    The key idea is to divide a long vapor cell into multiple smaller segments, and separate adjacent segments by a clear-air gap.
    This design allows the distribution of Rydberg atoms over an extended aperture for enhanced spatial signal reception without expanding the light attenuation distance.
    We develop a general signal model and analyze the resulting spatial reception pattern for this architecture.
    Our analysis demonstrates that, compared with the standard single-vapor-cell architecture, the segmental architecture effectively narrows the beamwidth. Moreover, the beamforming gain scales linearly with the number of segments in the BBR-limited regime, while its decay in the PSN-limited regime is substantially reduced. Consequently, the segmental design achieves significantly higher beamforming gains. We also analytically derive the optimal number of segments to maximize the beamforming gain.
\end{itemize}

\emph{Organization}: The remainder of the paper is organized as follows. Section~\ref{sec:2} presents the system model.  The analysis of atomic beamforming with a single vapor cell is provided in Section~\ref{sec:3} followed by beamforming analysis for the segmental vapor cells in Section~\ref{sec:4}. 
Numerical validations are conducted in Section~\ref{sec:5}, and conclusions are drawn in Section~\ref{sec:6}.

\emph{Notation}: Bold uppercase characters $\mb{X}$ denote matrices, with $X_{mn}$ representing its $(m,n)$-th entry. For two matrices $\mb{A}$ and $\mb{B}$, $[\mb{A}, \mb{B}]$ denotes the commutator $\mb{A}\mb{B}-\mb{B}\mb{A}$, and $\{\mb{A}, \mb{B}\}$ denotes their anti-commutator $\mb{A}\mb{B} + \mb{B}\mb{A}$. $\bra{p}$ and $\ket{q}$ refer to the bra-ket notation. $\mathcal{U}(a,b)$ denotes the uniform distribution over the interval $[a,b]$. Some default physical and mathematical constants are defined as follows: $j = \sqrt{-1}$ the imaginary unit; $c$ the speed of light; $h$ the Planck constant with $\hbar = \frac{h}{2\pi}$ its reduced form; $\epsilon_0$ the vacuum permittivity; $Z_0$ the vacuum impedance; $k_B$ the Boltzmann constant; and $q$ the elementary charge. 

\section{Models and Metrics} \label{sec:2}

\begin{figure*}
	\centering
	\includegraphics[width=6in]{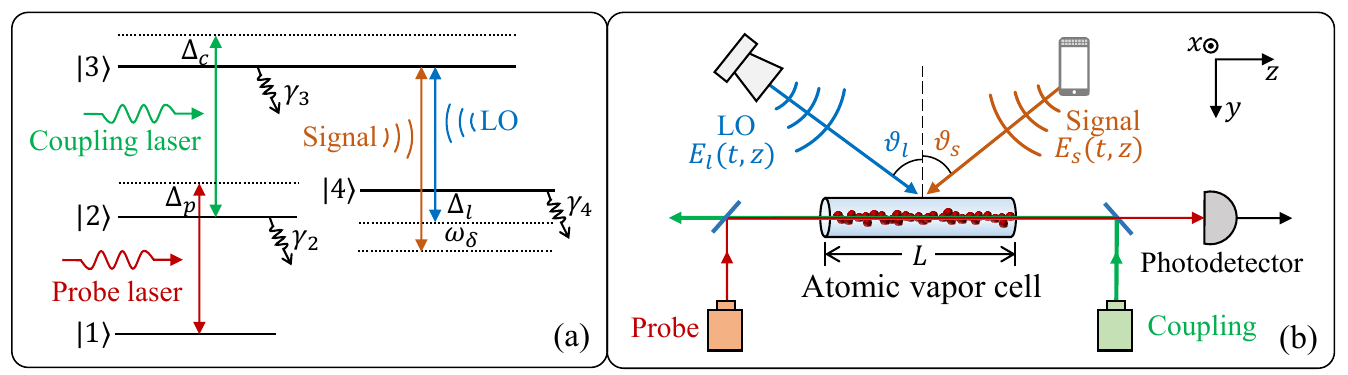}
	\vspace*{-1em}
	\caption{(a) Atomic energy levels and (b) architecture of a Rydberg atomic receiver
    .} 
	\vspace*{-1em}
	\label{img:system_model}
\end{figure*}

As illustrated in Fig.~\ref{img:system_model}, we consider an LO-dressed Rydberg superheterodyne receiver. The receiver consists of a pair of probe and coupling lasers that prepare the Rydberg atoms, together with a PD that measures the output probe-laser power.
A single vapor cell of length $L$, filled with alkali atoms, serves as the wireless signal sensor. 
This section presents the corresponding system models and performance metrics. The extension to the segmental-vapor-cell architecture is discussed in Section IV.

\subsection{Electric Field Models}
 A RARE detects the electric-field component of an incident electromagnetic wave. In the considered system, Rydberg atoms inside the vapor cell are exposed to three external fields: the LO field $E_l(t,z)$, the information-bearing signal field $E_s(t,z)$, and the BBR noise field $E_n(t,z)$. The total electric field at location $z\in[0, L]$ and time $t$ is given by the superposition of these three fields
\begin{align}
    E(t, z) = E_l(t,z) + E_s(t,z) + E_n(t,z).
\end{align}
For simplicity, we restrict the analysis to the field component polarized along the $x$-axis.

We follow the standard superheterodyne configuration~\cite{RydNP_Jing2020}. The LO field is slightly detuned from the signal field by a few kHz, thereby generating a beat frequency between the two fields. Let $\{f_{l}, \omega_{l}, \lambda_{l}, k_{l}\}$ and $\{f_{s}, \omega_{s}, \lambda_{s}, k_{s}\}$ denote the frequencies, angular frequencies, wavelengths, and wavenumbers of the LO and signal fields, respectively. The beat angular frequency is thus defined as $\omega_\delta\triangleq \omega_s - \omega_l$. 
To accurately characterize the effect of vapor-cell length on the quantum response, the spatial phase variations of the fields must be explicitly modeled.
Let $\vartheta_l$ and $\vartheta_s$ denote the angles-of-arrival of the LO and signal fields relative to the normal direction of the vapor cell, as shown in Fig.~\ref{img:system_model}(b). Under plane-wave propagation, the fields $E_l(t, z)$ and $E_s(t,z)$ are modeled as 
\begin{align}
    E_l(t, z) &= E_le^{j(\omega_lt- k_lz\theta_l + \phi_l)}, \label{eq:El}\\
    E_s(t,z)  &=  E_se^{j(\omega_s t - k_sz\theta_s + \phi_s)} \overset{(a)}{\approx} E_se^{j(\omega_s t - k_lz\theta_s + \phi_s)}. \label{eq:Es}
\end{align}
Here, $E_{l} > 0$, $\theta_{l}\triangleq\sin\vartheta_{l}$, $\phi_{l}$ (and $E_{s} > 0$, $\theta_{s}\triangleq\sin\vartheta_{s}$, $\phi_{s}$) represent the strengths, spatial directions, and phases of the LO (and signal) field. 
Approximation (a) holds since $\omega_\delta \ll \omega_l$. These field models explicitly capture both the temporal oscillation and the spatial phase profile of the incident waves.

For the BBR field $E_n(t,z)$, we adopt the isotropic propagation model~\cite{Noise_Luca2023, Noise_Wan2023}. In this model, the incident noise power is uniformly distributed over all spatial directions. Integrating the noise field over a unit spherical shell gives its autocorrelation function 
\begin{align}\label{eq:BBR_field}
    \mathbb{E}[E_n(t,z)E_n^*(t', z')] = \Lambda(f_l){\rm sinc}\left(\frac{2|z-z'|}{\lambda_l}\right)\delta(t\!-\!t').
\end{align}
The term $\Lambda(f_l) \triangleq  \frac{2\pi Z_0k_{B}T_{\rm env}f_l^2}{c^2}$ denotes the BBR spectral radiance at frequency $f_l$ and environmental temperature $T_{\rm env}$~\cite{Dynamic_Zhu2025}. The sinc term ${\rm sinc}\left(\frac{2|z-z'|}{\lambda_l}\right)$ characterizes the spatial correlation of the BBR noise, with ${\rm sinc}(u) = \frac{\sin(\pi u)}{\pi u}$. 
Since this work focuses on the spatial properties of atomic receivers, the temporal correlation is modeled by an impulse function $\delta(t-t')$ for simplicity. The analysis can be readily extended to more general temporal correlation models.

\subsection{Quantum System Model}

\subsubsection{Quantum state} As depicted in Fig.~\ref{img:system_model}(a), a RARE exploits Rydberg transitions to sense wireless signals.
Alkali atoms in the vapor cell are first excited from the ground state $\ket{1}$ to an intermediate state $\ket{2}$ by a probe laser of angular frequency $\omega_p$, and then to a Rydberg state $\ket{3}$ by the coupling laser at frequency $\omega_c$. These Rydberg atoms interact strongly with the incident field, $E(t,z)$, which induces the Rydberg transition $\ket{3}\rightarrow\ket{4}$. The quantum state of this four-level atomic system is described by a $4\times 4$ density matrix:
\begin{align}
    \boldsymbol{\rho} = \sum_{p=1}^4\sum_{q=1}^4\rho_{pq}\ket{p}\bra{q}\in\mathbb{C}^{4\times 4}.
\end{align}

\subsubsection{Quantum state evolution}
The evolution of the quantum state is governed by the Lindblad master equation~\cite{RydNP_Jing2020}: 
\begin{align}\label{eq:lindblad}
    \frac{\partial \boldsymbol{\rho}}{\partial t} = -\frac{j}{\hbar}[{\mb{H}}, \boldsymbol{\rho}] + \mathcal{L}. 
\end{align}
In this equation, the decoherence operator $\mathcal{L}$, expressed as 
    $\mathcal{L} = -\frac{1}{2}\{{\boldsymbol \Gamma}, \boldsymbol{\rho}\} + \boldsymbol{\Lambda}$,
where ${\boldsymbol \Gamma} = {\rm diag}\{0,\gamma_2,\gamma_3, \gamma_4\}$, $\boldsymbol{\Lambda} = {\rm diag}\{\gamma_2\rho_{22} + \gamma_4\rho_{44}, \gamma_3\rho_{33},0,0\}$, accounts for spontaneous decay among the atomic energy levels. Specifically, $\gamma_i$, $i=2,3,4$, denotes the spontaneous decay rates of the $i$-th level. 

By transforming the quantum system to a rotating frame defined by the laser and LO frequencies, the Hamiltonian operator $\mb{H}\in\mathbb{C}^{4\times 4}$ takes the form
\begin{align}\small\label{eq:hamiltonian}
    {\mb{H}} = \hbar\left[
    \begin{array}{cccc}
         0& \frac{\Omega_p}{2} &0 &0   \\
         \frac{\Omega_p}{2}& -\Delta_p & \frac{\Omega_c}{2} &0 \\
         0 & \frac{\Omega_c}{2} & -\Delta_p-\Delta_c & \frac{\Omega_l+\Omega_s^*+\Omega_n^*}{2} \\
         0 & 0 &  \frac{\Omega_l+\Omega_s+\Omega_n}{2} & -\Delta_p-\Delta_c-\Delta_l
    \end{array}
    \right].
\end{align}
Here,  $\Delta_p \triangleq \omega_p - \omega_{12}$, $\Delta_c \triangleq \omega_c - \omega_{23}$, and $\Delta_l \triangleq \omega_l - \omega_{34}$ are the frequency detunings from resonance, with $\omega_{pq}$ the transition frequency of $\ket{p}\rightarrow\ket{q}$. The quantities $\Omega_p$, $\Omega_c$, $\Omega_l$, $\Omega_{s}$, and $\Omega_{n}$ denote Rabi frequencies associated with the probe, coupling, LO, signal, and BBR fields, respectively. These Rabi frequencies characterize the intensity of the resonant electron transition. 
Because the Hamiltonian is represented in a rotating frame aligned with the probe, coupling, and LO fields,  $\Omega_p$, $\Omega_c$, and $\Omega_l$ are positive real-valued constants~~\cite{RydNP_Jing2020}. 
In contrast, $\Omega_{s}$ and $\Omega_{n}$ are complex functions of time $t$ and location $z$, reflecting the beat-frequency and spatial-phase variations. 
Taking the LO phase as a reference, we obtain~\cite{RydbergNoise_2024}
\begin{equation}   
    \begin{cases}
     \Omega_l  =  \frac{\mu_{34}}{\hbar}E_l  \\
     \Omega_{s}  =  \frac{\mu_{34}}{\hbar} E_s(t,z)e^{-j\angle E_l(t,z)}  \overset{(a)}{=}  \frac{\mu_{34}}{\hbar}E_se^{j(\omega_\delta t-k_lz\theta_\delta + \phi_\delta)} \\
     \Omega_{n}  =  \frac{\mu_{34}}{\hbar} E_n(t,z)e^{-j\angle E_{l}(t,z)}  \overset{(b)}{=}  \frac{\mu_{34}}{\hbar}E_n(t,z)e^{jk_lz\theta_l}
\end{cases},
\end{equation}
where $\mu_{34}$ is the transition dipole moment between $\ket{3}$ and $\ket{4}$. In (a), we define $\theta_\delta \triangleq\theta_s - \theta_l$ as the difference of spatial directions between LO and signal fields and $\phi_\delta \triangleq \phi_s-\phi_l$ as the phase difference. In (b), the phase $(\omega_l t + \phi_l)$ is absorbed in the noise $E_{n}(t,z)$. For later use, we define 
\begin{align}
    \Omega&(t,z) \triangleq |\Omega_l + \Omega_s(t,z) + \Omega_n(t,z)|
\end{align}
as the amplitude of the total Rabi frequency induced by the LO, signal, and BBR fields.

\subsubsection{Transmitted probe laser}
Inside the atomic medium, the transmission of the probe laser is governed by the quantum coherence $\rho_{12}$, 
the $(1,2)$-th entry of ${\boldsymbol{\rho}}$. Under the quasi-steady-state assumption, 
it is has been established that $\rho_{12}$ depends on the signal field $E_s(t,z)$ through the total Rabi-frequency amplitude~\cite{RydNP_Jing2020,Rydberg_Gong2025, liu_continuous-frequency_2022, Multiband_Cui2025}:
\begin{align}\label{eq:rho_12}
    \rho_{12} = \rho_{12}(\Omega(t,z)) = \rho_{12}(|\Omega_l + \Omega_s(t,z) + \Omega_n(t,z)|).
\end{align}
Our subsequent analysis relies only on this functional relationship, while the analytical expression of $\rho_{12}$ is not required. 

In addition, because of atomic thermal motion, the lasers illuminating atoms experience random Doppler shifts. Therefore, the observed quantum state should be averaged over the Doppler velocity distribution~\cite{Holloway2017ElectricFM}:
\begin{align}
    \bar{\rho}_{12} = \int_{-\infty}^{+\infty} \frac{e^{-{v^2}/{u^2}}}{\sqrt{\pi}u} \rho_{12}\left(\Delta_p - \frac{2\pi v}{\lambda_p}, \Delta_c + \frac{2\pi v}{\lambda_c}\right) {\rm d}v.
\end{align}
Here, $u \triangleq \sqrt{2k_B T_{\mathrm{env}} / m}$, where $m$ is the atomic mass, and the terms $\frac{2\pi v}{\lambda_p}$ and $\frac{2\pi v}{\lambda_c}$ denote the Doppler shifts associated with the probe and coupling lasers with an atomic velocity $v$.

Let $P_{\rm in}$ denote the probe-laser power at the input of the vapor cell. According to the Bouguer-Beer-Lambert law, the transmitted probe-laser power at the output of vapor cell is determined by the accumulated absorption along the cell
\begin{align}\label{eq:probe}
    P_{\rm out}(t) &= P_{\rm in}e^{-\int_0^L\chi(\Omega(t,z)){\rm d}z}.
\end{align} 
The power-attenuation coefficient, $\chi(\Omega(t,z))$, characterizes the susceptibility of the vapor cell to the probe laser at position $z$ and time $t$. This coefficient is determined by the Doppler-averaged quantum coherence by~\cite{RydNP_Jing2020}
\begin{align}
    \chi(\Omega(t,z)) = \frac{k_pN_0\mu_{12}^2}{\epsilon_0\hbar\Omega_p} \Im(\bar{\rho}_{12}), 
\end{align}
where $k_p$ is the probe-laser wavenumber, $N_0$ the atomic density, and $\mu_{12}$ the transition dipole moment from $\ket{1}$ to $\ket{2}$.

\subsubsection{ Quantum measurement model}
To perform quantum measurement, the Rydberg atomic receiver uses a PD to collect the transmitted probe-laser power and convert it into photocurrent~\cite{RydbergNoise_2024}:
\begin{align}
    I(t) &= \frac{q\eta}{\hbar\omega_p} P_{\rm out}(t) + \Delta I_{\rm psn}(t) \notag \\&= I_{\rm in} e^{-\int_0^L\chi(\Omega(t,z)){\rm d}z} + \Delta I_{\rm psn}(t),
\end{align}
Here, $\eta$ is the PD's quantum efficiency, $\Delta I_{\rm psn}(t)$ denotes the PSN, and $I_{\mathrm{in}} \triangleq \frac{q\eta P_{\mathrm{in}}}{\hbar\omega_p}$ is defined as the equivalent input current.

In practical deployments, the LO field is typically much stronger than the signal and BBR fields $\Omega_l \gg |\Omega_{s,n}(t,z)|$. 
Under this strong-LO condition, a first-order Taylor expansion yields a linearized decomposition of the photocurrent~\cite{QWS_Cui2025}:
\begin{align}\label{eq:decom}
    I(t) = \underbrace{\bar{I}}_{\rm Zero\:order} + \underbrace{\Delta I_{s}(t) + \Delta I_{\rm bbr}(t)}_{\rm First\:order} + \Delta I_{\rm psn}(t).  
\end{align}
 The zero-order current $\bar{I}$ is a time-invariant direct-current (DC) bias arising from the strong LO field $E_l(t,z)$; the oscillatory currents $\Delta I_{s}(t)$ and $\Delta I_{\rm bbr}(t)$ are associated with the information-carrying signal field $E_s(t,z)$ and the BBR noise $E_n(t,z)$, respectively. 
 The detailed derivation of this decomposition, together with explicit expressions for $\bar{I}$, $\Delta I_{s}(t)$, $\Delta I_{\rm bbr}(t)$, $ \Delta I_{\rm psn}(t)$ will be provided in the following sections. Without causing any confusion, we refer to $\Delta I_{s}(t)$ as ``signal" and $\Delta I_{\rm bbr}(t)$ as ``BBR noise" in the rest of the paper.

\subsection{Signal-to-Noise Ratio}
We use the SNR to evaluate the sensitivity of a RARE. Consider a time window of $T_s = \frac{2n\pi}{\omega_\delta}$, $n\in\mathbb{Z}^+$ for performing matched-filter detection. The window is assumed to be sufficiently long such that the spectra of the noises $\Delta I_{\rm bbr}(t)$ and $\Delta I_{\rm psn}(t)$ are approximately white within the bandwidth $B_s = \frac{1}{T_s}$. In this context, the SNR can be expressed as 
\begin{align}\label{eq:SNR0}
    {\rm SNR} = \frac{{P}_s}{{N}_{\rm bbr} + {N}_{\rm psn}}.
\end{align}
where $P_s$ denotes the accumulated signal energy:
 \begin{align}
     P_s = \int_0^{T_s} \Delta I_s^2(t){\rm d}t.
 \end{align}
The power spectral density of the BBR noise, $N_{\rm bbr}$, is obtained by integrating its correlation function over time:
 \begin{align}
     N_{\rm bbr} = \int_0^{T_s}R_{\rm bbr}(\tau){\rm d}\tau,
 \end{align}
 where $R_{\rm bbr}(\tau)\triangleq\mathbb{E}[\Delta I_{\rm bbr}(t)\Delta I_{\rm bbr}(t+\tau)]$ denotes the correlation function. 
Following the standard shot-noise model~\cite{Dynamic_Zhu2025}, the power spectral density of the photon shot noise $\Delta I_{\rm psn}(t)$ is proportional to the DC bias current.
\begin{align}
    N_{\rm psn}  =  {q}\bar{I}.
\end{align}

\subsection{A Myth and Reality}
\emph{1) Myth:}  Prior studies on RAREs have often adopted the point-receiver approximation when analyzing their reception characteristics~\cite{RydNP_Jing2020,Rydberg_Gong2025, liu_continuous-frequency_2022, Multiband_Cui2025, Dynamic_Zhu2025}. Under this approximation, the vapor-cell length $L$ is assumed to be sufficiently short so that the quantum state remains spatially invariant along the cell, i.e. $\bar{\rho}_{12}(\Omega(t,z))=\bar{\rho}_{12}(\Omega(t,0))$. This assumption simplifies the transmitted probe-laser power as $P_{\rm out}(t) = P_{\rm in}e^{-L\chi(\Omega(t,0))}$ and enables closed-form derivations of the signal energy $P_s$, total noise energy $N_{\rm bbr}+N_{\rm psn}$, and SNR. Based on this simplified model, RAREs have commonly been interpreted as \emph{isotropic} quantum antennas~\cite{RydMIMO_Sha2025, IsoRyd_Jia}.

\emph{2) Reality:} The point-receiver approximation, however, neglects the spatial variation of the quantum state $\bar{\rho}_{12}(\Omega(t,z))$ along the vapor cell.
This variation arises from the spatially dependent phase term $k_l z \theta_\delta$ imprinted on the Rabi frequency. 
When the vapor cell becomes sufficiently long, ignoring this spatial dependence leads to inaccurate modeling of the transmitted probe-laser power and the resulting SNR.  Consequently, the true reception characteristics of a practical Rydberg atomic receiver may be misinterpreted. The next section addresses this issue by explicitly accounting for the spatially varying quantum coherence.

\section{Atomic Beamforming with a Single Vapor Cell} \label{sec:3}
Building on the preceding “myth and reality” discussion, this section derives generalized signal and noise models that account for the spatial variation of the atomic quantum state. Our analysis shows that a RARE behaves as an isotropic quantum antenna only when the vapor cell is sufficiently short. By contrast, when the vapor cell is long, a directional receive beam naturally emerges, enabling atomic beamforming with a single vapor cell.

\subsection{ Mathematical Characterization of Signal, Noise, and SNR}
To establish comprehensive signal and noise models, we first derive a closed-form expression for the photocurrent decomposition in \eqref{eq:decom}. 
{\color{black} In practical systems, the LO field is usually orders of magnitude stronger than the signal field, i.e., $|E_{l}|\gg|E_{s}| $. This strong-LO condition arises because the LO source is placed close to the vapor cell, (e.g., 0.1$\sim$1 m), whereas the signal source is often tens to thousands of meters away. Due to wireless path loss, the incident signal power is typically $10^2$ to $10^6$ times weaker than that of the LO power when the two sources operate with comparable transmit powers. }
Under the strong-LO condition, the susceptibility can be accurately linearized around $\Omega_l$: 
\begin{align}\label{eq:chi}
   \chi(\Omega(t,z)) &\approx \chi(\Omega_l) + \frac{{\rm d}\chi}{{\rm d}\Omega_l}\Re[\Omega_s(t,z) + \Omega_n(t,z)]
    \notag \\&=\chi_l + \dot{\chi}_l\frac{\mu_{34}}{\hbar}\Re[E_s(t,z)e^{-j\angle E_l(t,z)}+ \notag \\ 
    &\quad\quad\quad\quad\quad\quad\quad\quad E_n(t,z)e^{-j\angle E_l(t,z)}].
\end{align}
The zero-order term $\chi_l \triangleq \chi(\Omega_l)$ and the first derivative $\dot{\chi}_l\triangleq\frac{{\rm d}\chi}{{\rm d}\Omega_l}$ depend only on the spatially invariant LO Rabi frequency $\Omega_l$. The first-order term captures the perturbations induced by the signal and BBR fields.
Substituting this linearized susceptibility into the probe-laser transmission model \eqref{eq:probe} gives the photocurrent decomposition in \eqref{eq:decom1}. Here, approximation (a) holds because $e^{-x}= 1 - x+\mathcal{O}(x^2)$. 
This decomposition enables us to separately characterize the time-varying signal current, $\Delta I_{s}$, the noises $\Delta I_{\rm bbr}$ and $\Delta I_{\rm psn}$, and their respective energies.  

\begin{figure*}[t!]
	\vspace*{-1em}
	\begin{align}\label{eq:decom1}
		I(t)  &= I_{\rm in}e^{-\int_{0}^L\chi(\Omega(t,z)){\rm d}z} + \Delta I_{\rm psn}(t) \approx  I_{\rm in}e^{-\int_{0}^L\chi_l{\rm d}z}
		e^{-\int_{0}^L\dot{\chi}_l\Re[\Omega_s(t,z) + \Omega_n(t,z)]{\rm d}z} + \Delta I_{\rm psn}(t)  \notag \\
		& \overset{(a)}{\approx} \underbrace{I_{\rm in}e^{-\chi_l L}}_{\bar{I}} \underbrace{-I_{\rm in}e^{-\chi_l L}\dot{\chi}_l\frac{\mu_{34}}{\hbar}\int_0^L\Re[E_s(t,z)e^{-j\angle E_{l}(t,z)}]{\rm d}z}_{\Delta I_s(t)} \underbrace{- I_{\rm in}e^{-\chi_l L}\dot{\chi}_l\int_0^L\Re[\Omega_n(t,z)]{\rm d}z}_{\Delta I_{\rm bbr}(t)} + \Delta I_{\rm psn}(t). 
	\end{align}  
		\hrulefill
\end{figure*}

\subsubsection{Signal characterization}
Evaluating the integral in \eqref{eq:decom1} yields the information‑bearing signal current:
\begin{align}\label{eq:signal}
   &\Delta I_{s}(t) = -I_{\rm in}e^{-\chi_l L}\dot{\chi}_l\frac{\mu_{34}}{\hbar}\int_0^L\Re[E_s(t,z)e^{-j\angle E_{l}(t,z)}]{\rm d}z\notag \\
   &=-\underbrace{I_{\rm in}e^{-\chi_l L}L{\rm sinc}\left(\frac{L\theta_\delta}{\lambda_l}\right)\dot{\chi}_l\frac{\mu_{34}}{\hbar}}_{{\rm Intrinsic\:gain\:}\kappa(\theta_\delta)}E_s\cos(\omega_\delta t + \phi_\delta'),
\end{align}
where $\phi_\delta' = \phi_\delta - \frac{\pi L\theta_\delta}{\lambda_l}$. 
The integral form in \eqref{eq:signal} reveals an essential insight: an atomic vapor cell functions as a \emph{virtual phased array} of length $L$, rather than merely as a single phase shifter as interpreted in earlier studies~\cite{RydGong_2025, RydNP_Jing2020}. 
 The received signal current is proportional to the integration of phase-shifted signal fields along the vapor cell, where the phase shift is determined by the spatial phase profile of the LO field $\angle E_l(t,z)$. Owing to the planar‑wavefront configuration, this integration naturally produces a directional dependence in the intrinsic gain  $\kappa(\theta_\delta)$ through the sinc function ${\rm sinc}\left(\frac{L\theta_\delta}{\lambda_l}\right)$.  As will be shown soon, this directional dependence is the key mechanism that distinguishes long vapor cells from short ones.

The signal energy is obtained by integrating $\Delta I_{s}^2(t)$ over the time window $T_s = \frac{2n\pi}{\omega_\delta},\forall n\in\mathbb{Z}^+$: 
\begin{align}\label{eq:Ps}
  P_s = \frac{1}{2}\kappa^2(\theta_\delta)E_s^2T_s. 
\end{align}

\subsubsection{Noise characterization}
Given the integral form of BBR noise $\Delta I_{\rm bbr}(t)$ in \eqref{eq:decom1}, its correlation function is derived as (see details in Appendix A)
    \begin{align}\label{eq:BBR1}
        R_{\rm bbr}(\tau) = \frac{1}{2}I_{\rm in}^2e^{-2\chi_l L}\xi\left(\frac{{L}}{\lambda_l};\theta_l\right)\lambda_l^2\dot{\chi}_l^2\frac{\mu_{34}^2}{\hbar^2}\Lambda(f_l)\delta(\tau),
    \end{align}
    where $\xi(d; \theta_l)\triangleq \int_{-d}^{d} (d-|u|)
    {\rm sinc}(2u)\cos{(2\pi\theta_lu)}{\rm d}u$.  The corresponding BBR noise power density is therefore
    \begin{align}\label{eq:N_BBR}
        N_{\rm bbr} = \frac{1}{2}I_{\rm in}^2e^{-2\chi_l L}\xi\left(\frac{{L}}{\lambda_l};\theta_l\right)\lambda_l^2\dot{\chi}_l^2\frac{\mu_{34}^2}{\hbar^2}\Lambda(f_l).
    \end{align}

For the photon shot noise, the DC bias $\bar{I} = I_{\rm in}e^{-\chi_l L}$ yields the PSN power density
\begin{align}\label{eq:N_PSN}
    N_{\rm psn} = q I_{\rm in}e^{-\chi_l L}. 
\end{align}

\subsubsection{SNR characterization} By combining \eqref{eq:Ps}–\eqref{eq:N_PSN}, the SNR at the photodetector is written as 
\begin{align}\label{eq:SNR1}
{\rm SNR} = \frac{{\rm sinc}^2\left(\frac{L\theta_\delta}{\lambda_l}\right)L^2e^{-\chi_l L}E_s^2T_s}{\xi\left(\frac{L}{\lambda_l};\theta_l\right)\lambda_l^2e^{-\chi_l L}\Lambda(f_l)+\beta_l}, 
\end{align}
where $\beta_l \triangleq \frac{2\hbar^2q}{I_{\rm in}\mu_{34}^2\dot{\chi}_l^2}$
is a constant that quantifies the relative strength of the photon shot noise.
Clearly, the SNR depends critically on the vapor-cell length $L$ and the spatial directions $\theta_l$ and $\theta_s$.   This dependence indicates that a single-vapor-cell RARE can exhibit a spatial reception pattern. In the following subsections, we analyze this pattern by examining the asymptotic SNR behavior in the short- and long-vapor-cell regimes.

\subsection{Short Vapor Cell as an Isotropic Quantum Antenna}\label{sec:3.2}
Existing studies mainly focus on the short-vapor-cell regime, where $L\ll\lambda_l$, and therefore often neglect the effect of cell length $L$ on the spatial response.
This subsection revisits this analysis and systematically discusses how $L$ influences the SNR, reception pattern, and effective aperture. 

When $L\ll\lambda_l$, the sinc term approaches unity, ${\rm sinc}\left({L\theta_\delta}/{\lambda_l}\right) \rightarrow 1$,  and  the BBR's correlation factor becomes
\begin{align}\label{eq:f1}
    \xi\left(\frac{L}{\lambda_l}; \theta_l\right) &\simeq  {\int_{-\frac{L}{\lambda_l}}^{\frac{L}{\lambda_l}} \left(\frac{L}{\lambda_l}-|u|\right) {\rm d}u}=\frac{L^2}{\lambda_l^2}. 
\end{align}
Substituting \eqref{eq:f1} into \eqref{eq:SNR1}, the asymptotic SNR is derived as 
\begin{align}\label{eq:SNR_small}
    {{\rm SNR} \simeq \frac{L^2e^{-\chi_l L}E_s^2T_s}{L^2e^{-\chi_l L}\Lambda(f_l)+\beta_l}}.
\end{align}
In this short-cell limit, the SNR becomes independent of the spatial directions $\theta_s$ and $\theta_l$. Consequently, the PD observes a uniform SNR regardless of the incident field directions, confirming that a short vapor cell behaves as an isotropic quantum antenna. 

A further observation is that both the signal-field intensity $E_s^2$ and the BBR-field strength $\Lambda(f_l)$ are scaled by a factor
\begin{align}
    \boxed{A_q {=} L^2e^{-\chi_l L}}, 
\end{align}
which has units of area.
This factor can be interpreted as the atomic aperture of the small quantum antenna, in analogy to the effective aperture of a classical antenna. Unlike its classical counterpart, the atomic aperture exhibits a tradeoff controlled by the cell length $L$. When the cell is too short, only a limited number of Rydberg atoms participate in the sensing process, which weakens the response to external electric fields. For longer vapor cells (yet still shorter than $\lambda_l$), the exponential attenuation of the probe laser also reduces the receiver sensitivity. Therefore, an intermediate cell length is required to maximize the atomic aperture.
By setting the derivative 
${\rm d}A_q/{\rm d}L$ to 0, the optimal cell length is given by 
\begin{align}\label{eq:L_opt}
    L^\star_{\rm short} = {2}/{\chi_l}.
\end{align}
The associated maximum atomic aperture in the short-cell regime is thereby inversely proportional to ${\chi_l^2}$:
\begin{align}
    A_q^\star = \frac{4}{e^2\chi_l^2}. 
\end{align}

In typical RARE implementations, the power attenuation coefficient often falls in the range $\chi_l\in [0.25,2]\:{\rm cm}^{-1}$. This corresponds to an optimal length $L^\star \in [1, 8]\:{\rm cm}$ and  a maximum atomic aperture $A_q^\star \in [0.135, 8.66]\:{\rm cm}^2$.
 It is important to note that the optimality condition in \eqref{eq:L_opt} holds only when $L \ll \lambda_l$. Since the maximum optimal length in this range is about $8\:\text{cm}$, the above aperture analysis applies primarily to frequency bands below roughly $3~\text{GHz}$.

\subsection{Long Vapor Cell as an Atomic Beamformer}
\begin{figure}
	\centering
	\includegraphics[width=3in]{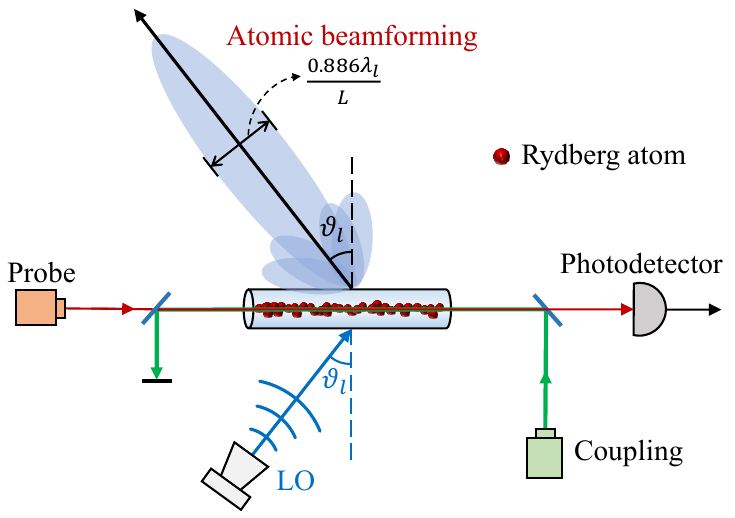}
	\vspace*{-1em}
	\caption{Long vapor cell as an atomic beamformer.} 
	\vspace*{-1em}
	\label{img:continuous_vc}
\end{figure}
We now turn to the long-cell regime, where $L \gg \lambda_l$.
This regime commonly arises at higher carrier frequencies, such as the midband (7-24 GHz) and mmWave band (30-300 GHz). In this regime, the sinc term in the SNR expression can no longer be ignored, and the power spectral density of BBR noise $N_{\rm bbr}$ must be re-examined.
\subsubsection{Asymptotic SNR}
Specifically, as $\frac{L}{\lambda_l}\to+\infty$,
the BBR's correlation factor $\xi(\frac{L}{\lambda_l};\theta_l)$ becomes asymptotically
\begin{align}\label{eq:f2}
    \xi\left(\frac{L}{\lambda_l}; \theta_l\right) &\simeq \frac{L}{\lambda_l} \int_{-\infty}^{\infty}{\rm sinc}(2u)\cos(2\pi\theta_l u){\rm d}u 
    \overset{(a)}{=}\frac{L}{2\lambda_l}.
\end{align}
Equality (a) follows directly from the Fourier transform of ${\rm sinc}(2u)$. Comparing \eqref{eq:f1} and \eqref{eq:f2}, we observe that the BBR correlation factor $\xi(\frac{L}{\lambda_l};\theta_l)$ shifts from a quadratic dependence on $\frac{L}{\lambda_l}$ to a linear dependence.
This reduction in scaling order occurs because BBR fields at widely separated positions along the cell become increasingly decorrelated. Consequently, the asymptotic SNR in the long-cell regime becomes
\begin{align}\label{eq:SNR_large}
    {{\rm SNR} \simeq {\rm sinc}^2\left(\frac{L\theta_\delta}{\lambda_l}\right)\frac{L^2e^{-\chi_l L}E_s^2T_s}{\frac{1}{2}L\lambda_le^{-\chi_l L}\Lambda(f_l)+\beta_l} } .
\end{align}

\subsubsection{\color{black}Receive beamforming pattern}
From \eqref{eq:SNR_large}, the long-cell SNR explicitly depends on the incident directions of the LO and signal fields. This directional dependence produces the receive-beam pattern of a long vapor cell:  
\begin{align}
    G(\theta_\delta) = \frac{P_s(\theta_\delta)}{P_{s,\max}} = {\rm sinc}^2\left(\frac{L\theta_\delta}{\lambda_l}\right),
\end{align} 
where $P_{s,\max}$ is attained at $\theta_{\delta}=0$.  A high SNR is therefore observed only when the signal direction $\theta_s$ closely aligns with the LO direction $\theta_l$, while a large angular deviation $\theta_\delta$ severely suppresses the signal and results in a low SNR. As illustrated in Fig.~\ref{img:continuous_vc},  this directional response forms a beam centered at $\theta_l$. Consequently, a long-cell RARE is transformed from an isotropic quantum antenna into an atomic beamformer.

The half-power beamwidth (HPBW), which quantifies the directivity, is obtained by solving the equation $G(\theta_\delta) = \frac{1}{2}$ as 
\begin{align}\label{eq:HPBW}
    {\theta_{\rm HPBW} = 0.886\frac{\lambda_l}{L}\:[\rm Rad]}.
\end{align}
Therefore, increasing the cell length or operating at a higher carrier frequency narrows the receive beam and improves angular selectivity.
\begin{remark}[Key features of an atomic beamformer]\rm
The atomic beamformer exhibits several distinctive characteristics:
    \begin{itemize}
    \item \textbf{Single-antenna beamforming capability}:
    The beamforming effect is achieved with a single quantum antenna consisting of one atomic vapor cell, one pair of laser beams, one PD, and one LO source. This configuration can be viewed as a quantum analogue of conventional analog beamforming: the PD acts as the RF chain, while the vapor cell functions as a continuous phased array whose phase shifts are engineered by the spatial profile of the LO field.
    \item \textbf{Low hardware cost}:  Vapor cells are substantially more cost‑effective than traditional phased arrays of comparable aperture. At higher frequencies, conventional arrays require a linearly increasing number of antenna elements to preserve aperture, accompanied by complex RF circuitry~\cite{Phaseshifter_Roi2016}.
    In contrast, the atomic vapor cell intrinsically provides a continuous aperture due to the $\mu$m-level atomic spacing. Moreover, a single vapor cell can operate over a broad range of frequency bands without hardware modification by exploiting the rich energy-level structure of Rydberg atoms.
    \item \textbf{Beam direction control}:  The beamforming direction is determined by the incident angle of the LO field, as shown in Fig.~\ref{img:continuous_vc}.  To adaptively steer the beam, the LO source can be mounted on a motor‑driven circular rail, enabling mechanical adjustment of the LO direction $\theta_l$. 
\end{itemize}
\end{remark}

\subsubsection{Atomic beamforming gain} 
  To better understand the fundamental limits of atomic beamforming gain, we analyze the SNR scaling law under two noise regimes.
  The first is the BBR-limited regime, ${\rm SNR}_{\rm bbr} = \frac{P_s}{N_{\rm bbr}}$, where the receiver is dominated by environmental noise. The second is the PSN-limited regime, ${\rm SNR}_{\rm psn} = \frac{P_s}{N_{\rm psn}}$, where photon shot noise dominates. The overall SNR is the harmonic mean of the SNRs associated with these two regimes: 
 \begin{align}
 {\rm SNR} = \frac{{\rm SNR}_{\rm bbr}{\rm SNR}_{\rm psn}}{{\rm SNR}_{\rm bbr} + {\rm SNR}_{\rm psn}}.
 \end{align}

\emph{i) Beamforming gain for aligned $\theta_s$ and $\theta_l$}:  
Considering perfect alignment between the LO and signal directions, we have $\theta_\delta = 0$ and ${\rm sinc}^2\left(\frac{L\theta_\delta}{\lambda_l}\right) = 1$. The SNR expressions in the BBR- and PSN-limited regimes then simplify to
\begin{align}\label{eq:SNR_bbr_psn1}
    \left\{ 
    \begin{array}{ll}
       {\rm SNR}_{\rm bbr} \simeq \frac{2LE_s^2T_s}{\lambda_l\Lambda(f_l)} = \mathcal{O}\left(\frac{L}{\lambda_l}\right)  \\
       {\rm SNR}_{\rm psn} \simeq \frac{L^2e^{-\chi_l L}E_s^2T_s}{\beta_l} = \mathcal{O}\left(L^2e^{-\chi_l L}\right) 
    \end{array}
    \right..
\end{align}
In the BBR-limited regime, the beamforming gain scales linearly with $\frac{L}{\lambda_l}$, which is analogous to the array gain of a classical phased array.
This scaling arises because the signal fields add coherently along the vapor cell, whereas the BBR fields add incoherently when the cell length is much larger than the wavelength. Hence, a longer cell can more effectively suppress external BBR noise. 
In the PSN‑limited regime,  the SNR scales with the atomic aperture  $A_q = L^2e^{-\chi_lL}$. This relationship comes because the signal intensity  $E_s^2$ is amplified by $L^2e^{-2\chi_lL}$, while the PSN power $N_{\rm psn}$ scales with the DC bias $\bar{I}\propto e^{-\chi_l L}$. 
Their ratio precisely gives the atomic aperture.

\begin{remark}[Tradeoff between the BBR- and PSN-limited regimes]\rm
     Although the BBR-limited regime suggests that the beamforming gain can grow without bound as the cell length increases, the PSN-limited regime imposes a fundamental limitation. As $L\rightarrow+\infty$, although ${\rm SNR}_{\rm bbr}$ improves linearly,  both the signal energy $P_s$ and the BBR's power density $N_{\rm bbr}$ decay exponentially faster than the PSN's power density $N_{\rm psn}$. As a result, the system eventually enters the PSN-limited regime, and the overall SNR experiences an exponential roll-off. This tradeoff implies the existence of an optimal vapor-cell length that maximizes the SNR. Therefore, proper cell-length selection is essential for realizing the full beamforming gain of a single vapor cell.
\end{remark}

\emph{ii) Beamforming gain outside the HPBW}: 
We next examine the beamforming gain for signals arriving outside the main lobe, i.e., for $|\theta_\delta|>\frac{1}{2}\theta_{\rm HPBW}$. Using the bound ${\rm sinc}^2\left(x\right)\le \frac{1}{\pi^2 x^2}$, the SNRs in the two regimes are upper-bounded by
\begin{align}
\left\{ 
    \begin{array}{l}
       {\rm SNR}_{\rm bbr} \lesssim \frac{2\lambda_lE_s^2T_s}{\pi^2\theta_\delta^2L\Lambda(f_l)} =   \mathcal{O}\left(\frac{\lambda_l}{L\theta_\delta^2}\right)  \\
       {\rm SNR}_{\rm psn} \lesssim \frac{\lambda_l^2e^{-\chi_l L}E_s^2T_s}{\pi^2\theta_\delta^2\beta_l} =  \mathcal{O}\left(\frac{\lambda_l^2}{\theta_\delta^2e^{\chi_l L}}\right)
    \end{array}
    \right. .
\end{align}
In both regimes, the upper bounds decay quadratically with the angular deviation $\theta_{\delta}$. As the cell length 
$L$ increases, the BBR-limited SNR falls linearly, while the PSN-limited SNR drops exponentially. These results confirm that a longer vapor cell not only narrows the beamwidth but also actively suppresses signals outside the main lobe. This inherent spatial filtering capability enables effective interference mitigation and can support spatial‑division multiple access in quantum‑enhanced wireless systems.

{\subsection{Optimal Vapor Cell Length} 
In this subsection, we derive the optimal cell length $L$ for a long vapor cell. As discussed above, increasing the cell length narrows the beamwidth, but it does not necessarily improve the beamforming gain in the PSN-limited regime.
To balance directivity and beamforming gain, we find the optimal length $L$  that maximizes the SNR while satisfying a given angular-selectivity constraint. 

Specifically, the optimization problem is formulated as
\begin{align}
    \max_{L>0} \quad&\frac{L^2e^{-\chi_l L}E_s^2T_s}{\frac{1}{2}L\lambda_le^{-\chi_l L}\Lambda(f_l)+\beta_l}, \label{eq:P0}\\
     {\rm s.t.} \quad&0.886\frac{\lambda_l}{L}\le\theta_\epsilon,
     \tag{\ref{eq:P0}a}\label{eq:P0a}
\end{align}
where \eqref{eq:P0a} enforces that the HPBW is no wider than a prescribed threshold $\theta_{\epsilon}$. The globally optimal solution to \eqref{eq:P0} is (see Appendix B for the derivation)
\begin{align}\label{eq:optimalL}
    L^\star = \max\left\{\frac{1}{\chi_l} \left[2 + F^{-1}\left(\frac{\Lambda(f_l)\lambda_l}{2\chi_l\beta_l}\right)\right], 0.886\frac{\lambda_l}{\theta_\epsilon} \right\},
\end{align}
where 
$F\left(x\right) := \frac{xe^{x+2}}{x+2}$ and $F^{-1}$ denotes its inverse. The first term, $L_{\rm long}^\star = \frac{1}{\chi_l} \left[2 + F^{-1}\left(\frac{\Lambda(f_l)\lambda_l}{2\chi_l\beta_l}\right)\right]$, represents the unconstrained length that maximizes the SNR,  whereas the second term, $0.886\frac{\lambda_l}{\theta_\epsilon}$ guarantees the desired directivity.

\begin{figure}
	\centering
	\includegraphics[width=3.3in]{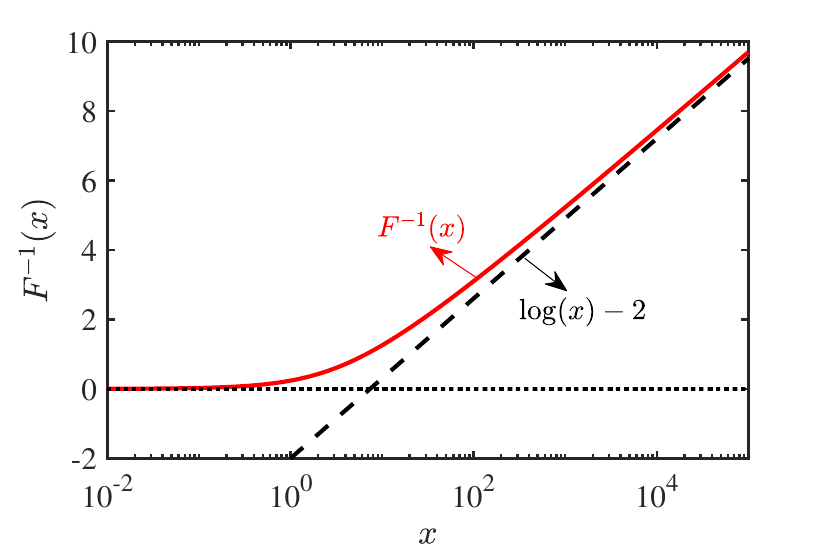}
	\vspace*{-1em}
	\caption{The function $F^{-1}(x)$ together with its small‑ and large‑argument asymptotics.} 
	\vspace*{-1em}
	\label{img:Fx_inv}
\end{figure}

To gain further insight, we examine the unconstrained optimum $L_{\rm long}^\star$. Compared with the optimal short‑cell length $L_{\rm short}^\star = \frac{2}{\chi_l}$,  the long‑cell optimum  $L_{\rm long}^\star$ contains an additional term  $\frac{1}{\chi_l} F^{-1}\left(\frac{\Lambda(f_l)\lambda_l}{2\chi_l\beta_l}\right)$. 
Fig.~\ref{img:Fx_inv} plots $F^{-1}(x)$, which possesses the following useful properties: 
1) $F^{-1}(x) > 0$ and  is strictly increasing for $x > 0$; 2) for large $x$,  $F^{-1}(x)\simeq \log(x)-2$.  These properties allow us to analyze the key features of $L_{\rm long}^\star$. 
\begin{remark}[Key features of the optimal cell length]
Since $F^{-1}\left(\frac{\Lambda(f_l)\lambda_l}{2\chi_l\beta_l}\right) > 0$,  the optimal length in the long‑cell regime always exceeds that in the short‑cell regime. 
The argument $\frac{\Lambda(f_l)\lambda_l}{2\chi_l\beta_l}$ effectively measures the relative strength of BBR noise compared with PSN. 
When BBR noise dominates, a longer cell is preferred to achieve higher beamforming gain. 
In the extreme BBR-dominated regime  $\frac{\Lambda(f_l)\lambda_l}{2\chi_l\beta_l}\rightarrow +\infty$, the optimal long-cell length scales logarithmically with the BBR-to-PSN strength ratio:
\begin{align}\label{eq:asmp}
	L_{\rm long}^\star \simeq \frac{1}{\chi_l}\log\left(\frac{\Lambda(f_l)\lambda_l}{2\chi_l\beta_l}\right).
\end{align}
\end{remark}

In practice, the BBR field always dominates. For example, using the representative parameters adopted in the numerical validation section, namely $f_l = 6.9458\:{\rm GHz}$, $\chi_l = 42.4\:{\rm m}^{-1}$, and $\dot{\chi}_l = 2.08\times10^{-5}\:{\rm m}^{-1}{\rm Hz}^{-1}$, the BBR-to-PSN strength ratio is $\frac{\Lambda(f_l)\lambda_l}{2\chi_l\beta_l} = 3198.5$. In this context,  the logarithmic approximation in \eqref{eq:asmp} is highly accurate, and the optimal cell length is enlarged by $\frac{L_{\rm long}^\star}{L_{\rm short}^\star}\approx\frac{\log\left(\frac{\Lambda(f_l)\lambda_l}{2\chi_l\beta_l}\right)}{2} \approx 4$ times relative to the short-cell case. 
}

\begin{figure}
	\centering
	\includegraphics[width=3in]{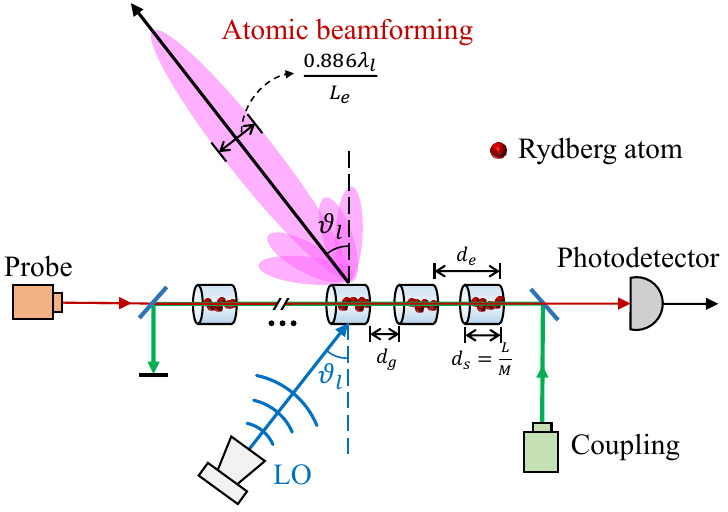}
	\vspace*{-1em}
	\caption{Architecture of segmental vapor cells.} 
	\vspace*{-1em}
	\label{img:seg_vc}
\end{figure}

\begin{figure*}[b!]\color{black}
		\vspace*{-1em}
		\hrulefill
	\begin{align}\label{eq:decom2}
		&I(t)  = I_{\rm in}n_t^{2M}e^{-\sum\limits_{m=0}^{M-1}\int_{md_e}^{md_e+d_s}\chi(\Omega(t,z)){\rm d}z} + \Delta I_{\rm psn}(t) \approx  I_{\rm in}n_t^{2M}e^{-\chi_l L}
		e^{-\sum\limits_{m=0}^{M-1}\int_{md_e}^{md_e + d_s}\dot{\chi}_l\Re[\Omega_s(t,z) + \Omega_n(t,z)]{\rm d}z} + \Delta I_{\rm psn}(t)
		\notag \\
		&\approx \underbrace{I_{\rm in}n_t^{2M}e^{-\chi_l L}}_{\bar{I}} \underbrace{-I_{\rm in}n_t^{2M}e^{-\chi_l L}\dot{\chi}_l\sum_{m = 0}^{M-1}\int_{md_e}^{md_e+d_s}\Re[\Omega_s(t,z)]{\rm d}z}_{\Delta I_s(t)}  \underbrace{- I_{\rm in}n_t^{2M}e^{-\chi_l L}\dot{\chi}_l\sum_{m = 0}^{M-1}\int_{md_e}^{md_e+d_s}\Re[\Omega_n(t,z)]{\rm d}z}_{\Delta I_{\rm bbr}(t)} + \Delta I_{\rm psn}(t). 
	\end{align}
\end{figure*}

\section{Atomic Beamforming with Segmental Vapor Cells} \label{sec:4}
Although atomic beamforming with a single vapor cell can narrow the beamwidth by increasing the cell length $L$, the beamforming gain does not grow  monotonically. 
Instead, the gain eventually decreases in the PSN regime due to the exponential attenuation of laser power inside the atomic medium. 
This limitation constrains the achievable beamforming performance of the single-cell architecture.
To mitigate this issue, we propose a \emph{segmental-vapor-cell} architecture, as illustrated in Fig.~\ref{img:seg_vc}. 
The key idea is to uniformly divide a long vapor cell of length $L$ into $M$ shorter segments. These segments are arranged collinearly along the optical propagation path, with a clear-air gap $d_g$ inserted between each pair of adjacent segments. 
By increasing either the number of segments 
$M$ or the gap $d_g$, the effective reception area can be enlarged without increasing the total light-attenuation distance $L$. As a result, this architecture can achieve a narrower beamwidth and higher beamforming gain than the single-vapor‑cell design. {\color{black} Notably, the segmental-vapor-cell architecture still requires only a single pair of probe and coupling lasers to probe the atom-field interaction, together with a single PD to convert the transmitted probe light into photocurrent.}

\subsection{Mathematical Characterization of Signal, Noise, and SNR}

We now derive the signal and noise models for the segmental-cell architecture.  A single pair of laser beams sequentially  traverses all $M$ cell segments, and the output probe light is collected by one PD.
Let the physical length of each segment be $d_s = \frac{L}{M}$ and the center‑to‑center spacing between adjacent segments be $d_e = d_s + d_g$. 
The total effective aperture occupied by all segments is thus $L_e = L + (M-1)d_g$. 
{\color{black}Compared with the single-cell architecture, the output probe power in the segmental architecture must account for both the absorption of the probe laser inside each segment and the reflection losses at the $2M$ cell‑air interfaces, which is thus expressed as 
\begin{align} 
  P_{\rm out}(t) &= P_{\rm in} n_t^{2M}e^{-\sum\limits_{m = 0}^{M-1}\int_{md_e}^{md_e+d_s}\chi(\Omega(t,z)){\rm d}z}.
\end{align}
Here, $n_t \in [0, 1]$ denotes the power-transmission coefficient at each interface.} In practice, $n_t$ could approach 1 using anti‑reflection coatings, e.g., 
commercial coatings readily achieve $n_t \ge 0.995$~\cite{ARcoating}. 

Applying the strong-LO linearization in \eqref{eq:chi}, the photocurrent measured at the PD can be decomposed as shown in \eqref{eq:decom2}. Based on this decomposition, we next characterize the received signal $\Delta I_s(t)$, noises $\Delta I_{\rm bbr}(t)$ and $\Delta I_{\rm psn}(t)$, and the resulting SNR.

\subsubsection{Signal characterization} 
Evaluating the sum of integrals in \eqref{eq:decom2} over all vapor-cell segments gives the signal component of the photocurrent:
\begin{align}\label{eq:sig2}
    \Delta I_s(t) &= -\underbrace{I_{\rm in}n_t^{2M}e^{-\chi_l L}L{\rm sinc}\left(\frac{d_s\theta_\delta}{\lambda_l}\right)\Xi_M\left(\frac{d_e\theta_\delta}{\lambda_l}\right) \dot{\chi}_l\frac{\mu_{34}}{\hbar}}_{{\rm Intrinsic\:gain\:\kappa(\theta_\delta)}}
\notag\\&\quad\quad\quad\quad\quad \times E_s\cos(\omega_\delta t + \phi_\delta''),
\end{align}
where $\Xi_M(x)\triangleq\frac{\sin(M\pi x)}{M\sin(\pi x)}$ is the Dirichlet sinc function, and $\phi_\delta''= \phi_\delta - \frac{\pi d_s\theta_\delta}{\lambda_l} - \frac{\pi(M-1) d_e\theta_\delta}{\lambda_l}$. 
The accumulated signal energy over the matched-filter observation window $T_s=\frac{2n\pi}{\omega_{\delta}},\forall n\in\mathbb{Z}^+$
is
\begin{align} \label{eq:Ps2}
   P_s &= \frac{1}{2}\kappa^2(\theta_\delta)E_s^2T_s.
\end{align}
Equations \eqref{eq:sig2} and \eqref{eq:Ps2} show that the reception pattern of segmental vapor cells is governed by the product of ${\rm sinc}^2\left(\frac{d_s\theta_\delta}{\lambda_l}\right)$ and $\Xi_M^2\left(\frac{d_e\theta_\delta}{\lambda_l}\right)$:
\begin{align}
    G(\theta_\delta) =  \frac{P_s(\theta_\delta)}{P_{s,\max}} = {\rm sinc}^2\left(\frac{d_s\theta_\delta}{\lambda_l}\right) \Xi_M^2\left(\frac{d_e\theta_\delta}{\lambda_l}\right).
\end{align} 
This pattern follows the \emph{principle of pattern multiplication} in classical antenna theory~\cite{antennabook}. Specifically, the sinc term arises from the continuous integration of the electromagnetic field within each vapor-cell segment and therefore characterizes the element pattern of an individual segment. The Dirichlet sinc term arises from the discrete summation of the fields across all segments and therefore characterizes the array factor of the segmental vapor-cell structure.

Similar to a long single vapor cell, the segmental vapor cells form a receive beam steered toward the LO direction $\theta_s = \theta_l$ (see Fig.~\ref{img:seg_vc}). The HPBW in this configuration is primarily determined  by the array factor $\Xi_M^2\left(\frac{d_e\theta_\delta}{\lambda_l}\right)$, because the effective aperture  $L_e = L + (M - 1)d_g \approx L + Md_g = Md_e$ is much larger than one segment's length $d_s = \frac{L}{M}$ when $M$ is large. Solving $\Xi_M^2\left(\frac{d_e\theta_\delta}{\lambda_l}\right) = \frac{1}{2}$ with $\theta_{\rm HPBW} = 2\theta_\delta$ thereby yields the HPBW: 
\begin{align}\label{eq:HPBW2}
{\theta_{\rm HPBW} =  0.886\frac{\lambda_l}{ Md_e} \approx 0.886 \frac{\lambda_l}{L_e}\:[{\rm Rad}]}.
\end{align}
\begin{remark}[Narrower beamwidth]\rm
Comparing \eqref{eq:HPBW} and \eqref{eq:HPBW2}, we see that the HPBW of segmental vapor cells is no longer determined solely by the physical atomic-medium length $L$, but by the effective aperture $L_e = L + (M-1)d_g$.
Owing to the introduced clear-air gap 
$d_g$, the segmental cells produce a narrower beam than a single cell of the same physical length 
$L$. Furthermore, the beamwidth can be reduced by either increasing the number of segments 
$M$ or enlarging the gap $d_g$, without extending the physical length 
$L$.  
\end{remark}

\subsubsection{Noise characterization} 

We next derive the BBR noise power density for the segmental architecture. To generate a narrow receive beam, this work considers a relatively large clear-air gap $d_g$, typically comparable to the wavelength. In this regime, the BBR fields impinging on different vapor-cell segments are weakly correlated, and their cross-correlations can be safely neglected. This assumption is consistent with the standard modeling approach used in conventional antenna array theory.
Under this condition, the BBR correlation function is derived as (see Appendix C for details)
\begin{align}
    R_{\rm bbr}(\tau) = 
    \frac{1}{2}I_{\rm in}^2Mn_t^{4M}e^{-2\chi_l L}\xi\left(\frac{d_s}{\lambda_l};\theta_l\right)\lambda_l^2\dot{\chi}_l^2\frac{\mu^2_{34}}{\hbar^2}\Lambda(f_l)\delta(\tau),
\end{align}
which results in the BBR noise power density:
\begin{align}\label{eq:BBR2}
  N_{\rm bbr}=  \frac{1}{2}I_{\rm in}^2Mn_t^{4M}e^{-2\chi_l L}\xi\left(\frac{d_s}{\lambda_l};\theta_l\right)\lambda_l^2\dot{\chi}_l^2\frac{\mu^2_{34}}{\hbar^2}\Lambda(f_l). 
\end{align}

For photon shot noise, the DC bias, $\bar{I} = I_{\rm in} n_t^{2M} e^{-\chi_l L}$, is now determined by both the laser absorption inside atomic medium and reflection losses on the cell surfaces. Thus, the corresponding PSN power density becomes
\begin{align}
	N_{\rm psn} = q I_{\rm in}e^{-\chi_l L}n_t^{2M}. 
\end{align}

\subsubsection{SNR characterization} Combining the accumulated signal energy and the noise power densities, the SNR observed at the PD is formulated as 
\begin{align}\label{eq:SNR2}
{\rm SNR} = \frac{{\rm sinc}^2\left(\frac{d_s\theta_\delta}{\lambda_l}\right)\Xi_M^2\left(\frac{d_e\theta_\delta}{\lambda_l}\right)n_t^{2M}L^2e^{-\chi_l L}E_s^2T_s}{Mn_t^{2M}\xi\left(\frac{d_s}{\lambda_l};\theta_l\right)\lambda_l^2e^{-\chi_l L}\Lambda(f_l)+\beta_l}. 
\end{align}
In the following subsections, we analyze the achievable beamforming gain of segmental vapor cells by examining the asymptotic SNRs in both the short-segment and long-segment regimes.

\subsection{\color{black} Beamforming Gain with Short Cell Segments}

\subsubsection{Asymptotic SNR} We first consider the regime where each vapor-cell segment is much shorter than the wavelength, $d_s\ll \lambda_l$. As established in the short-cell analysis in Section~\ref{sec:3.2}, the BBR correlation factor $\xi\left(\frac{d_s}{\lambda_l};\theta_l\right)$ is asymptotically 
$\xi(\frac{d_s}{\lambda_l};\theta_l) \simeq \frac{d_s^2}{\lambda_l^2} = \frac{L^2}{M^2\lambda_l^2}$. Substituting this into \eqref{eq:SNR2} yields the asymptotic SNR for the short-segment regime:
\begin{align}\label{eq:SNR_small2}
{
{\rm SNR} \simeq \frac{{\rm sinc}^2\left(\frac{d_s\theta_\delta}{\lambda_l}\right)\Xi_M^2\left(\frac{d_e\theta_\delta}{\lambda_l}\right)n_t^{2M}L^2e^{-\chi_l L}E_s^2T_s}{\frac{1}{M}n_t^{2M}L^2e^{-\chi_l L}\Lambda(f_l)+\beta_l}}. 
\end{align}

\subsubsection{Atomic beamforming gain} 
From \eqref{eq:SNR_small2}, we can examine the beamforming gain in the BBR- and PSN-limited regimes, respectively. Considering perfect alignment between the LO and signal fields,  the SNRs in the two regimes simplify to
\begin{align}\label{eq:SNR_cs_small}
    \begin{cases}
       {\rm SNR}_{\rm bbr} \simeq \frac{ME_s^2T_s}{\Lambda(f_l)} = \mathcal{O}\left(M\right)  \\
       {\rm SNR}_{\rm psn} \simeq \frac{n_t^{2M}L^2e^{-\chi_l L}E_s^2T_s}{\beta_l} = \mathcal{O}\left(n_t^{2M}L^2e^{-\chi_l L}\right) 
    \end{cases}.
\end{align}

{\color{black}
Several important insights can be drawn from \eqref{eq:SNR_cs_small}. First, because BBR fields in different cell segments are uncorrelated, the BBR‑noise power density is reduced by a factor  $\frac{1}{M}$ relative to a short single vapor cell.  Therefore, the segmental architecture achieves a beamforming gain that scales linearly with $M$ in the BBR-limited regime. 
Second, the segmental architecture introduces reflection losses at the $2M$ cell interfaces. Thereby, in the PSN-limited regime, the SNR contains an additional factor $n_t^{2M}$ in addition to the single-vapor-cell atomic-aperture factor, $A_q$. This factor captures the cumulative optical loss caused by interface reflections.

Compared with the single-cell design, the segmental architecture enables the enhancement of beamforming gain by adjusting the number of segments $M$, rather than relying solely on increasing the physical cell length $L$.
Although the PSN‑limited SNR also decays exponentially with $M$, this penalty is minimal when high‑quality anti‑reflection coatings (with a transmission coefficient $n_t$ approaching 1) are employed. For example, with 
$M=50$ and a transmission coefficient of $n_t=0.995$, the reflection loss contributes only about 2.17~dB to the PSN‑limited SNR. When $n_t=0.999$, the loss is reduced to only about 0.43~dB.  \emph{Thus, the optical attenuation caused by interface reflection can be made much weaker than the attenuation caused by the atomic medium, enabling the segmental-vapor-cell architecture to achieve a substantially higher SNR than the single-vapor-cell counterpart.} This result will be confirmed numerically in the simulation section.
}

\subsection{\color{black}Beamforming Gain with Long Cell Segments}
\subsubsection{Asymptotic SNR}
We next consider the regime where each vapor-cell segment is long compared with the wavelength,  i.e.,
$d_s \gg \lambda_l$. This case differs from the short-segment regime mainly in the BBR noise power density. As shown earlier, when $\frac{d_s}{\lambda_l}\to+\infty$,  we have 
$\xi\left(\frac{d_s}{\lambda_l};\theta_l\right) \simeq \frac{d_s}{2\lambda_l}$. Substituting this into \eqref{eq:SNR2} yields the asymptotic SNR in the long-segment regime:
\begin{align}\label{eq:SNR_large2}
{{\rm SNR} \simeq \frac{{\rm sinc}^2\left(\frac{d_s\theta_\delta}{\lambda_l}\right)\Xi_M^2\left(\frac{d_e\theta_\delta}{\lambda_l}\right)n_t^{2M}L^2e^{-\chi_l L}E_s^2T_s}{\frac{1}{2}n_t^{2M}L\lambda_le^{-\chi_l L}\Lambda(f_l)+\beta_l}}. 
\end{align}

\subsubsection{Atomic beamforming gain} 
When the signal direction is aligned with the LO direction, the SNRs in the BBR-limited and PSN-limited regimes simplify to
\begin{align}\label{eq:SNR_bbr_psn2}
    \left\{ 
    \begin{array}{ll}
       {\rm SNR}_{\rm bbr} \simeq \frac{2LE_s^2T_s}{\lambda_l\Lambda(f_l)} = \mathcal{O}\left(\frac{L}{\lambda_l}\right)  \\
       {\rm SNR}_{\rm psn} \simeq \frac{n_t^{2M}L^2e^{-\chi_l L}E_s^2T_s}{\beta_l} = \mathcal{O}\left(n_t^{2M}L^2e^{-\chi_l L}\right) 
    \end{array}
    \right..
\end{align}
It is noteworthy that the BBR‑limited SNR for long segments is identical to that of a long single vapor cell.
This equivalence arises because each segment is sufficiently long, so that the BBR field within it exhibits spatial correlation comparable to that in a single cell of length 
$L$. In contrast, the PSN-limited SNR suffers from the additional factor $n_t^{2M}$ due to reflection losses at the $2M$ interfaces. 
Therefore, by comparing \eqref{eq:SNR_bbr_psn1} and \eqref{eq:SNR_bbr_psn2}, we see that the SNR achieved by segmental vapor cells with long segments is always lower than that of a single vapor cell of the same physical length $L$. \emph{This observation indicates that long segments are not beneficial for the segmental architecture. To obtain a higher beamforming gain, the segment length should be made as short as possible so that the system operates in the short-segment regime.}

\subsection{The Optimal Number of Segments}

\begin{figure}
	\centering
	\includegraphics[width=3.3in]{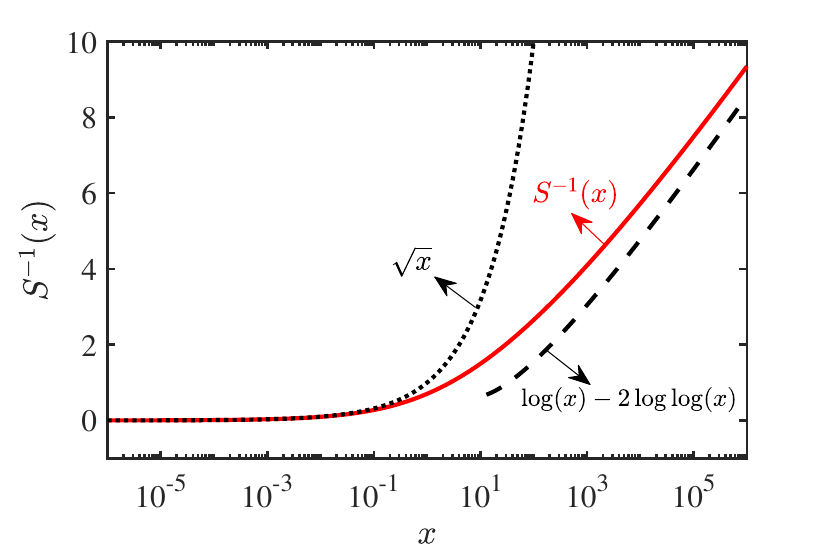}
	\vspace*{-1em}
	\caption{The function $S^{-1}(x)$ together with its small‑ and large‑argument asymptotics.} 
	\vspace*{-1em}
	\label{img:Sx_inv}
\end{figure}

In this subsection, we optimize the number of segments $M$ to achieve the highest beamforming gain. Since the long-segment regime is not advantageous, as established above, we focus on the short-segment regime. 
Moreover, because the HPBW can be independently controlled by adjusting the clear-air gap $d_g$, the number of segments can be optimized solely for beamforming gain maximization. 
This leads to the following optimization problem:
\begin{align}\label{eq:P1}
\max_{M>0} \frac{n_t^{2M}L^2e^{-\chi_l L}E_s^2T_s}{\frac{1}{M}n_t^{2M}L^2e^{-\chi_l L}\Lambda(f_l)+\beta_l},
\end{align}
where, for analytical convenience, $M$ is treated as a continuous variable. The resulting continuous optimum will subsequently be quantized to the nearest positive integer for practical implementation. 
The globally optimal solution to \eqref{eq:P1} is (see Appendix D for the derivation)
\begin{align}\label{eq:optimalM}
    M^\star = \frac{1}{2|\log n_t|} S^{-1} \left(\frac{2|\log n_t| L^2e^{-\chi_lL}\Lambda(f_l)}{\beta_l}\right),
\end{align}
where $S(x) := x^2e^x$ and $S^{-1}$ denotes its inverse. Fig.~\ref{img:Sx_inv} plots $S^{-1}(x)$, which exhibits the following useful properties: 
1) $S^{-1}(x) > 0$ and  is strictly increasing for $x > 0$; 2) for small $x$, $S^{-1}(x)\simeq \sqrt{x}$; 3)    for large $x$,  $S^{-1}(x)\simeq \log(x)-2\log\log(x)$. 
These properties allow us to characterize the optimal number of segments in the limits of high and low interface transmission coefficients.
\begin{remark}[Optimal $M^\star$ for high and low transmission coefficients] When high-quality anti-reflection coatings are used,  $n_t\to 1$ and $|\log n_t|\to 0$. In this limit, the optimal number of segments scales as 
	\begin{align} \label{eq:Mn1}
		M^\star \simeq \sqrt{\frac{ L^2e^{-\chi_lL}\Lambda(f_l)}{2|\log n_t|\beta_l}} = \mathcal{O}\left(\frac{1}{\sqrt{|\log n_t|}}\right).
	\end{align}
This trend is intuitive: as the interface transmission improves, reflection loss becomes weaker, which raises the PSN-limited SNR and allows the receiver to use more segments to further suppress external BBR noise.

In the opposite limit where  $n_t \rightarrow 0$ (and hence $|\log n_t|\rightarrow +\infty$), we have
\begin{align} \label{eq:Mn2}
	M^\star = \mathcal{O}\left(\frac{\log|\log n_t|}{|\log n_t|}\right). 
\end{align}
Equation \eqref{eq:Mn2} indicates that the number of segments decreases as $n_t$ becomes smaller. This behavior is expected because reflection loss becomes the dominant factor limiting beamforming gain in this regime.  Thereby, fewer segments are recommended to minimize the cumulative reflection penalty. 
\end{remark}

\section{Numerical Validation} \label{sec:5}
\begin{figure*}
	\centering
	\includegraphics[width = 1\textwidth]{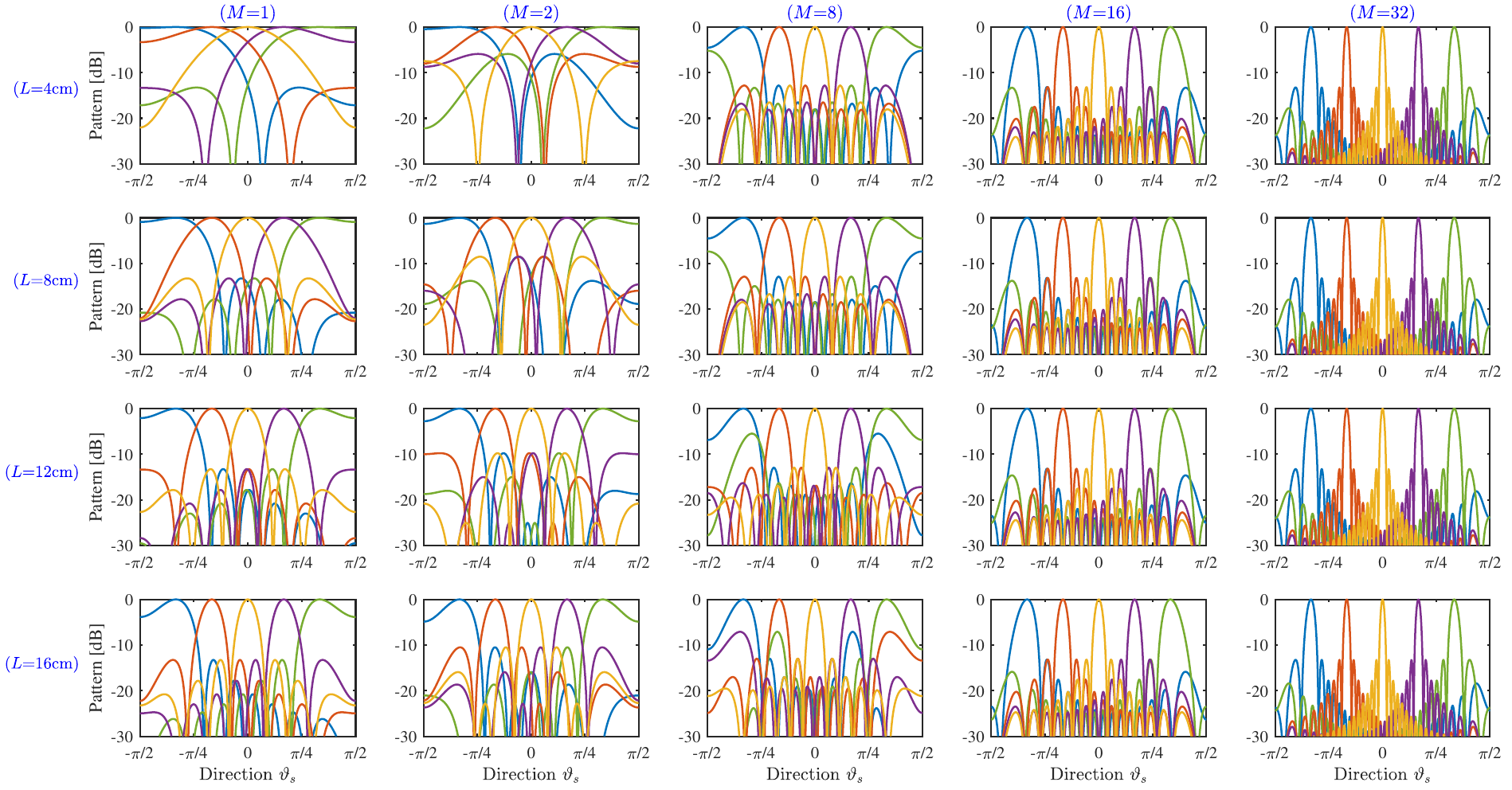}
	\vspace*{-1em}
	\caption{Normalized beamforming patterns achieved by the single- and segmental-vapor-cell architectures with varying vapor-cell lengths and segment numbers. 
		Rows correspond to $L = \{4,8,12,16\}$ cm, and columns correspond to $M = \{1,2,8,16,32\}$. 
	} 
	\vspace*{-1em}
	\label{img:spatial_pattern1}
\end{figure*}

\subsection{Simulation Setup}
In this section, we numerically validate the discovered atomic beamforming phenomena. 
Unless otherwise specified, the following default parameters are used. The Rydberg atomic receiver employs a cesium‑133 vapor cell with atomic density $N_0 = 4.89\times 10^{10}\:{\rm cm}^{-3}$. The four relevant energy levels are $6S_{1/2}$, $6P_{3/2}$, $47D_{5/2}$, and $48P_{3/2}$; the corresponding decay rates are taken from~\cite{RydNP_Jing2020}. 
The vapor cell is illuminated by a probe laser at 852~nm with a Rabi frequency of $\Omega_p = 2\pi\times5.7\:${MHz} (input power $P_{\rm in} = 120\:{\rm \mu W}$), and a coupling laser at~509 nm with a Rabi frequency of $2\pi\times 0.89\:${MHz} (input power of $40\:{\rm mW}$). The LO source is placed 1~meter from the vapor cell center. It delivers an LO field at $f_l = 6.9458\:{\rm GHz}$ with a transmit power of $-17$~dBm, which yields an incident field strength $E_l = 34.6\:{\rm mV/m}$.  The corresponding wavelength is $\lambda_l = 4.32\:{\rm cm}$.   
Based on these parameters, the susceptibility and its derivative are numerically computed as $\chi_l = 42.4\:{\rm m}^{-1}$ and $\dot{\chi}_l = 2.08\times10^{-5}\:{\rm m}^{-1}{\rm Hz}^{-1}$. 
The signal field strength is set to $E_s = 154.9\:{\rm \mu V/m}$.  
The frequency offset and time window are configured as $\omega_\delta = 2\pi \times 100\:{\rm kHz}$ and $T_s = 10\:{\rm \mu s}$, respectively. All noise sources are exposed to an environmental temperature at $T_{\rm env} = 290\:{\rm K}$, and the photodetector's quantum efficiency is $\eta = 0.8$. 

\subsection{Simulation Results}

{\color{black} Fig.~\ref{img:spatial_pattern1}  presents the normalized beamforming patterns for both single- and segmental-vapor-cell architectures. 
Four cell lengths ($L \in \{4,8,12,16\}\:{\rm cm}$)  and five segment numbers ($M \in \{1,2,8,16,32\}$)  are examined. 
The case $M = 1$ exactly corresponds to a standard single-vapor-cell configuration. 
For $M>1$, the clear-air gap is chosen as $d_g = \max\{\frac{\lambda_l}{2} - \frac{L}{M}, \frac{\lambda_l}{4}\}$, ensuring that the gap never falls below $\frac{\lambda_l}{4} = 1.08\:{\rm cm}$. 
Each sub-figure illustrates the pattern at five LO directions:  $-\frac{\pi}{3}$ (blue line), $-\frac{\pi}{6}$ (orange line),  $0$ (yellow line), $\frac{\pi}{6}$ (purple line), and $\frac{\pi}{3}$ (green line), while the signal direction is swept from $-\frac{\pi}{2}$ to $\frac{\pi}{2}$. 
We first consider the single-vapor-cell results, which correspond to the first column with $M = 1$. The beam is steered towards the direction aligned with the LO, indicating that beam steering can be accomplished by mechanically adjusting the position and/or orientation of the LO source. 
As the cell length increases from $4\:{\rm cm}$ to $16\:{\rm cm}$, the HPBW narrows approximately from $54.8^\circ$ to $13.7^{\circ}$.  A sharp angular selectivity is obtained at a long vapor cell. This selectivity can be further enhanced by adopting the segmental architecture.  For a fixed cell length, increasing the number of segments from 
1 to 32 progressively reduces the beamwidth.
For example, with $L = 4$~cm, the HPBW narrows from $54.8^\circ$ (single cell) to about $3.2^\circ$ (32 segments); with $L = 16$~cm, the HPBW similarly decreases from $13.7^\circ$ to $3.2^\circ$. Notably, when 
$M$ is sufficiently large, the beamwidth becomes almost independent of the original cell length $L$. This behavior arises because the overall pattern is dominated by the array factor  $\Xi_M^2\left(\frac{d_e\theta_\delta}{\lambda_l}\right)$. This result validates the capability  of the proposed segmental architecture in enhancing angular selectivity. }

\begin{figure}
	\centering
	\includegraphics[width = 0.49\textwidth]{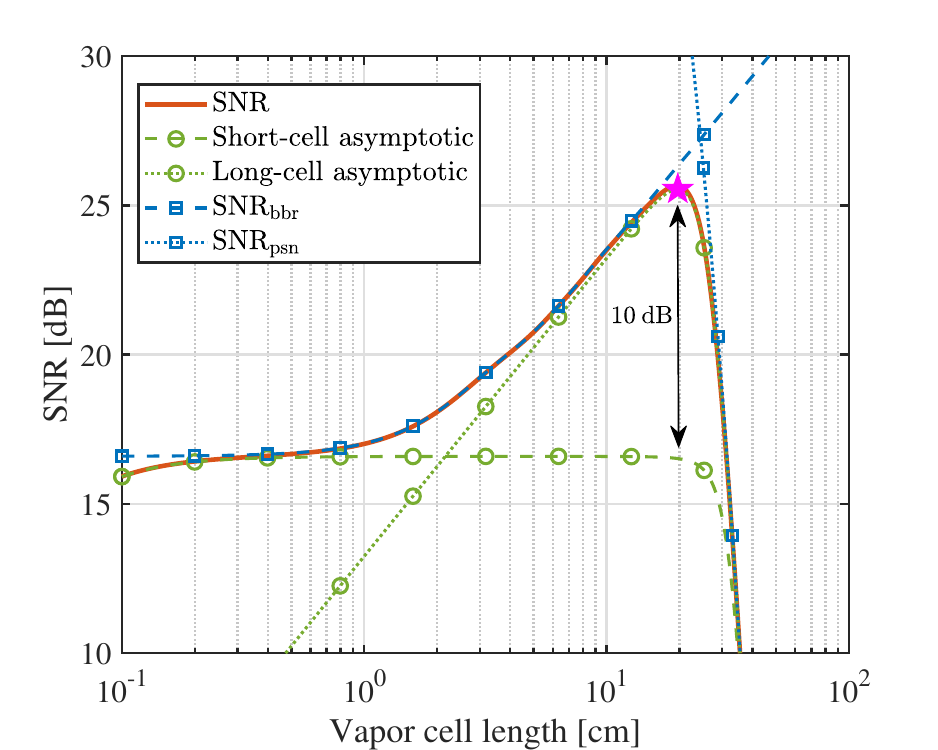}
	\vspace*{-1em}
	\caption{Influence of length $L$ of a single vapor cell on SNR. The pink star represents the maximum SNR achieved by the optimal unconstrained cell length.} 
	\vspace*{-1em}
	\label{img:SNR_L}
\end{figure}

Fig.~\ref{img:SNR_L} illustrates the influence of cell length $L$ on the beamforming gain for a single vapor cell. The length spans from 0.1~cm to 100~cm, and the beam peak is aligned to the signal direction. Five SNR curves are plotted, including the exact SNR from \eqref{eq:SNR1}, the short-cell asymptotic in \eqref{eq:SNR_small}, the long-cell asymptotic in \eqref{eq:SNR_large}, and BBR- and PSN-limited approximations from \eqref{eq:SNR_bbr_psn1}. The pink star  marks the maximum SNR achieved by the optimal unconstrained length $L_{\rm long}^\star = \frac{1}{\chi_l} \left[2 + F^{-1}\left(\frac{\Lambda(f_l)\lambda_l}{2\chi_l\beta_l}\right)\right] = 21\:$cm.
As observed, the short- and long-cell asymptotic expressions accurately capture the SNR scaling in their respective regimes, consistent with the analytical derivations.
The two asymptotic curves intersect near $L = 2\:{\rm cm}$. 
More importantly, the SNR initially increases with $L$ and then falls rapidly. This behavior clearly reflects the tradeoff between the BBR- and PSN-limited regimes. In the BBR-limited region, extending $L$ from 1~cm to 21~cm effectively improves the total SNR by 10~dB. This is attributed to the $\mathcal{O}(L)$ beamforming gain that suppresses external BBR noise. Beyond $L = 21\:{\rm cm}$, however, the PSN regime becomes dominant, and the SNR decays exponentially owing to the severe laser attenuation. The maximum SNR occurs precisely at the analytically derived optimal length $L_{\rm long}^\star$, confirming its value as a practical design guideline for atomic vapor cells.

\begin{figure}
	\centering
	\includegraphics[width = 0.49\textwidth]{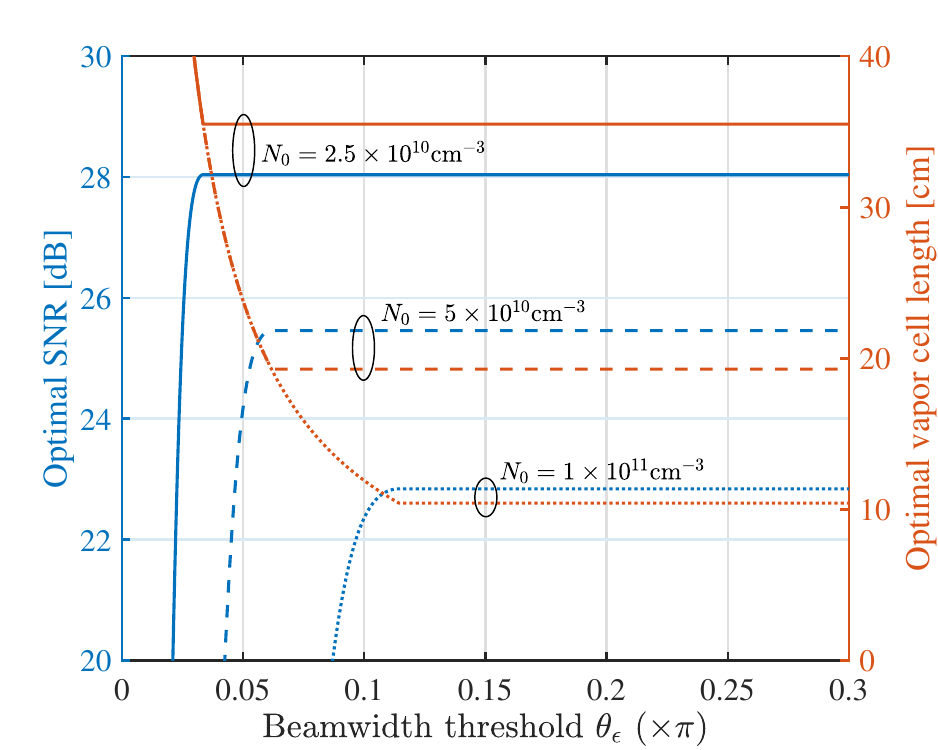}
	\vspace*{-1em}
	\caption{Influence of beamwidth threshold $\theta_\epsilon$ on the optimal cell length $L^\star$ and the achieved SNR.} 
	\vspace*{-1em}
	\label{img:SNR_bw}
\end{figure}

 Fig.~\ref{img:SNR_bw} investigates the impact of the beamwidth threshold $\theta_\epsilon$ on both the optimal cell length $L^\star$ under an angular selectivity constraint and the corresponding achievable SNR. Three atomic densities are compared: $N_0 \in \{2.5,5,10\}\times 10^{10}\:{\rm cm}^{-3}$. As the beamwidth threshold decreases from $\pi/2$ to $0$, the angular selectivity requirement becomes increasingly stringent. In this scenario, the optimal cell length $L^\star$ initially remains at the unconstrained optimum $L_{\rm long}^\star$, since this length already satisfies the target angular selectivity. As $\theta_\epsilon$ decreases further, $L^\star$ begins to scale as $0.886\lambda_l/\theta_\epsilon$. This increase in cell length reduces the maximum achievable SNR. To limit the SNR degradation to within 3~dB, the cell length must be kept below 49, 26, and 14$\:$cm for $N_0 = 2.5\times 10^{10}, 5\times 10^{10}$, and $1\times10^{11}\:{\rm cm}^{-3}$, which results in the minimum achievable beamwidth of $0.025\pi$, $0.046\pi$, and $0.086\pi$. 

\begin{figure}
	\centering
	\includegraphics[width = 0.49\textwidth]{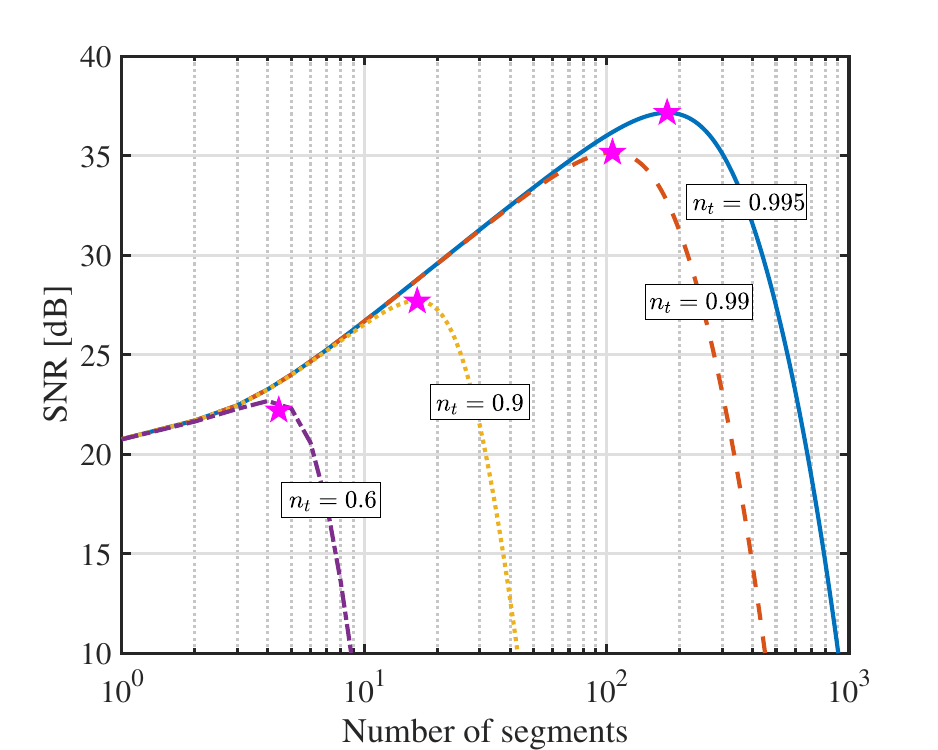}
	\vspace*{-1em}
	\caption{Achieved SNR as a function of the number of segments $M$ and the power-transmission coefficient $n_t$. The pink stars represent the maximum SNR obtained at $M^\star$.} 
	\vspace*{-1em}
	\label{img:SNR_nt}
\end{figure}

In Fig.~\ref{img:SNR_nt}, we investigate the achievable beamforming gain of segmental vapor cells.
The SNR, as defined in \eqref{eq:SNR2}, is evaluated for different numbers of segments $M\in[1, 1000]$ and for four transmission coefficients $n_t\in\{0.6, 0.9, 0.99, 0.995\}$. 
In this simulation, the cell length $L$ is fixed at 5~cm and the clear-air gap is set to $d_g = \max\{\frac{\lambda_l}{2} - \frac{L}{M}, \frac{\lambda_l}{4}\}$. The pink star in each curve marks the maximum SNR achieved at the analytically optimal number of segments $M^\star$. For all considered transmission coefficients, the SNR initially increases and then declines. This behavior stems from the tradeoff between the $\mathcal{O}(M)$-scaling BBR-limited SNR and the $\mathcal{O}(n_t^{2M})$-decaying PSN-limited SNR. Nevertheless, adopting the optimal number of segments consistently yields an SNR exceeding that of the single-vapor-cell configuration ($M=1$). Moreover, the maximum achievable SNR increases substantially with higher transmission coefficients. For $n_t = 0.9$, using $M^\star = 17$ segments can improve the SNR of a single vapor cell by 7~dB, while for $n_t = 0.995$, employing $M^\star = 178$ segments yields a higher enhancement of 17~dB. These results highlight the importance of high-quality anti-reflection coatings for realizing the benefits of the segmental architecture. Notably, commercial anti-reflection coated vapor cells can achieve transmission coefficients exceeding 0.995 at the relevant optical wavelengths (852~nm and 509~nm)~\cite{ARcoating}, firmly supporting the performance advantages of the proposed segmental architecture.

\begin{figure}
	\centering
	\includegraphics[width = 0.49\textwidth]{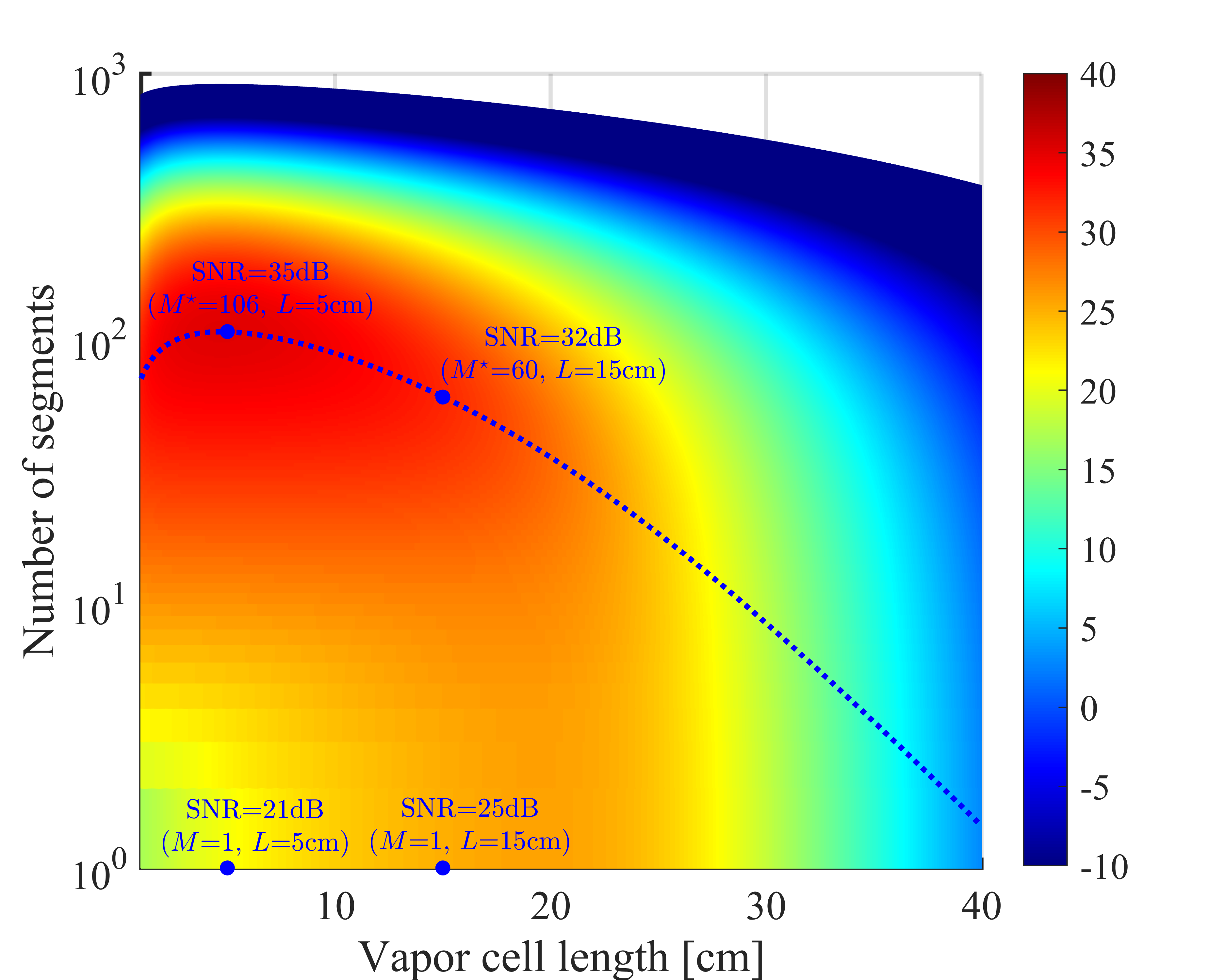}
	\vspace*{-1em}
	\caption{Achieved SNR at different numbers of segments $M$ and cell lengths $L$. The dashed blue curve depicts the optimal number of segments $M^\star$ for each cell length.} 
	\vspace*{-1em}
	\label{img:SNR_ML}
\end{figure}

{\color{black}Fig.~\ref{img:SNR_ML} further examines the SNR enhancement capability of segmental vapor cells across varying  cell lengths. In this study, $M$ varies from 1 to 1000, $L$ varies from 1~cm to 40~cm, and the resulting SNRs are recorded. The dashed blue curve represents the optimal number of segments $M^\star$ for each cell length, with the transmission coefficient fixed at $n_t = 0.99$. 
For all cell lengths considered, judiciously selecting the number of segments consistently improves the SNR relative to the standard single-vapor-cell architecture ($M = 1$). For instance, at $L = 5$~cm, the SNR improves from 21~dB to 35~dB when the number of segments is increased from 1 to 106; at $L = 15$~cm, the SNR increases from 25~dB to 32~dB as $M$ grows from 1 to 60. Conversely, when the number of segments becomes excessively large (e.g., $M > 500$), the SNR degrades significantly due to the reflection loss. These findings validate that the proposed segmental architecture offers a substantial enhancement in beamforming gain and highlight the importance of optimizing the number of segments.}

\begin{figure*}[b!]
\vspace*{-1em}
\hrulefill	
    \begin{align}\label{ap:1}
    R_{\rm bbr}(\tau) &= \left(I_{\rm in}e^{-\chi_l L}\dot{\chi}_l\frac{\mu_{34}}{\hbar}\right)^2\int_0^L\int_0^L
    \mathbb{E}\{\Re[E_n(t,z)e^{jk_lz\theta_l}]\Re[E_n( t+ \tau,z')e^{jk_lz'\theta_l}]\}{\rm d}z {\rm d}z'\notag\\
    &= \frac{1}{2}\left(I_{\rm in}e^{-\chi_l L}\dot{\chi}_l\frac{\mu_{34}}{\hbar}\right)^2\int_0^L\int_0^L
    \Re\{\mathbb{E}[E_n(t,z)E_n^*(t+ \tau,z')]e^{jk_l(z-z')\theta_l}\}{\rm d}z {\rm d}z'\notag \\
    &\overset{(a)}{=} \frac{1}{2}\left(I_{\rm in}e^{-\chi_l L}\dot{\chi}_l\frac{\mu_{34}}{\hbar}\right)^2\Lambda(f_l)\delta(\tau)\int_0^L\int_0^L
    {\rm sinc}(2(z-z')/\lambda_l)\cos{(2\pi(z-z')\theta_l/\lambda_l)}{\rm d}z {\rm d}z'\notag \\
    & \overset{(b)}{=} \frac{1}{2}\left(I_{\rm in}e^{-\chi_l L}\dot{\chi}_l\frac{\mu_{34}}{\hbar}\right)^2\Lambda(f_l)\delta(\tau)\int_{-L}^L (L-|u|)
    {\rm sinc}(2u/\lambda_l)\cos{(2\pi\theta_lu/\lambda_l)} {\rm d}u\notag\\
    & \overset{(c)}{=} \frac{1}{2}I_{\rm in}^2e^{-2\chi_l L}\xi\left(\frac{L}{\lambda_l};\theta_l\right)\lambda_l^2\dot{\chi}_l^2\frac{\mu^2_{34}}{\hbar^2}\Lambda(f_l)\delta(\tau).
\end{align}
\end{figure*}

\begin{figure*}[t!]
\begin{align}\label{ap:2}
    R_{\rm bbr}(\tau) &= \left(I_{\rm in}e^{-\chi_l L}n_t^{2M}\dot{\chi}_l\frac{\mu_{34}}{\hbar}\right)^2\sum_{m,m'}\int_{md_e}^{md_e+d_s}\int_{m'd_e}^{m'd_e+d_s}
    \mathbb{E}\{\Re[E_n(t,z)e^{jk_lz\theta_l}]\Re[E_n(t+\tau,z')e^{jk_lz'\theta_l}]\}{\rm d}z {\rm d}z'\notag\\
    &\overset{(a)}{=} \sum_{m}\frac{1}{2}\left(I_{\rm in}e^{-\chi_l L}n_t^{2M}\dot{\chi}_l\frac{\mu_{34}}{\hbar}\right)^2\int_{md_e}^{md_e+d_s}\int_{md_e}^{md_e+d_s}
    \Re\{\mathbb{E}[E_n(t,z)E_n^*(t+\tau,z')]e^{jk_l(z-z')\theta_l}\}{\rm d}z {\rm d}z'\notag \\
    & = \sum_{m}\frac{1}{2}\left(I_{\rm in}e^{-\chi_l L}n_t^{2M}\dot{\chi}_l\frac{\mu_{34}}{\hbar}\right)^2\Lambda(f_l)\lambda_l^2\int_{-d_s/\lambda_l}^{d_s/\lambda_l} (d_s/\lambda_l-|u|)
    {\rm sinc}(2u)\cos{(2\pi\theta_lu)}\delta(\tau){\rm d}u  \notag \\
    & = \frac{1}{2}I_{\rm in}^2e^{-2\chi_l L}n_t^{4M}M\xi\left(\frac{d_s}{\lambda_l};\theta_l\right)\lambda_l^2\dot{\chi}_l^2\frac{\mu^2_{34}}{\hbar^2}\Lambda(f_l)\delta(\tau).
\end{align}
\hrulefill
\vspace*{-1em}
\end{figure*}

\section{Conclusions} \label{sec:6}
This paper presented a theoretical analysis of the beamforming behavior of Rydberg atomic receivers based on single- and segmental-vapor-cell architectures.
We demonstrated that a long vapor cell exhibits a directional reception pattern with a beamwidth inversely proportional to cell length, revealing that the vapor cell functions as a continuous virtual phased array. 
We also analytically derived the optimal cell length that maximizes the beamforming gain, which captures the fundamental tradeoff between BBR- and PSN-limited regimes. 
To mitigate PSN-induced gain decay, we proposed a segmental-vapor-cell architecture that extends the effective reception aperture without increasing optical attenuation distance. Numerical results demonstrated a 10$\:$dB SNR enhancement through optimal cell length selection, and an additional 17$\:$dB improvement enabled by the segmental architecture.

Despite these contributions, the theoretical framework established in this paper marks a starting point for Rydberg atomic beamforming, and several important issues remain open for future investigation. For example,  the current approach relies on mechanically rotating the LO source to steer the beam direction. In practical systems, digital beam-direction control is crucial for supporting mobile users and adapting to dynamic propagation environments. Another promising direction is to investigate the beamforming behavior in a multi-vapor-cell system equipped with multiple laser beams and photodetectors, which could serve as a quantum analogue of conventional hybrid analog-digital combining. 
In addition, experimental validation of the predicted atomic beamforming phenomenon remains essential to establish its practical viability. 
{\color{black}Finally, the segmental-vapor-cell architecture might encounter several hardware imperfections, such as segment-to-segment variations in temperature and atomic densities.
	These imperfections may transform the segmental vapor cells from a uniformly weighted array into a non-uniformly weighted one.
	In practical implementations, such imperfections should be carefully addressed through engineering approaches, such as temperature stabilization.
	Moreover, analyzing the beamforming pattern and gain under these non-idealities is also an important direction for future work. 
}


\appendix
\subsection{Proof of \eqref{eq:BBR1}}
Adopting the form of $\Delta I_{\rm bbr}(t)$ in \eqref{eq:decom1}, the correlation function $R_{\rm bbr}(\tau)$ is derived in \eqref{ap:1}. Here, (a) arises from the correlation function of the BBR field in \eqref{eq:BBR_field}, (b) follows from the change of integral variables $u = z - z'$ and $v = z + z'$, and (c) comes from the definition $\xi(d; \theta_l)\triangleq \int_{-d}^{d} (d-|u|) {\rm sinc}(2u)\cos{(2\pi\theta_lu)}{\rm d}u$. This completes the proof. 
\subsection{Proof of \eqref{eq:optimalL}}
We begin by deriving the optimal solution to \eqref{eq:P0} without considering the beamwidth constraint. Through algebraic manipulation, the objective function can be expressed in normalized form as
\begin{align}
    f(\bar{L}) = \frac{\bar{L}^2}{a\bar{L}+e^{\bar{L}}},
\end{align}
where $\bar{L}:= \chi_lL$ and $a:= \frac{\lambda_l\Lambda(f_l)}{2\beta_l\chi_l}$. Maximizing the SNR is therefore equivalent to maximizing $f(\bar{L})$. The derivative of $f(\bar{L})$ is
\begin{align}
    \frac{{\rm d} f(\bar{L})}{{\rm d}\bar{L}} = \frac{\bar{L}}{(a\bar{L}+e^{\bar{L}})^2}\underbrace{\left[a\bar{L}-(\bar{L} - 2)e^{\bar{L}}\right]}_{f_1(\bar{L})}.
\end{align}
We examine the auxiliary function $f_1(\bar{L})$. Its derivative $f_1'(\bar{L}) = a - (\bar{L} - 1)e^{\bar{L}}$ shows that $f_1(\bar{L})$ increases for small $\bar{L}$ and  eventually decreases as $\bar{L}$ grows, with $f_1(0) = 0$ and $\lim_{\bar{L}\rightarrow +\infty} f_1(\bar{L}) = -\infty$. Hence, $f_1(\bar{L})$ has exactly one positive root, denoted as $\bar{L}^\star$. Setting $f_1(\bar{L}^\star) = 0$ yields
\begin{align}
a\bar{L}^\star = (\bar{L}^\star - 2)e^{\bar{L}^\star} \Rightarrow    \bar{L}^\star = 2+F^{-1}(a), 
\end{align}
where $F(x):=\frac{xe^{x+2}}{x+2}$. 
Because $f_1(\bar{L}) > 0$ for $\bar{L}<\bar{L}^\star$ and $f_1(\bar{L}) < 0$ for $\bar{L}>\bar{L}^\star$, the derivative $\frac{{\rm d} f(\bar{L})}{{\rm d}\bar{L}}$ changes sign only at $\bar{L}^\star$. Thus $f(\bar{L})$ is unimodal and attains its maximum at $\bar{L}^\star$. Reverting to the original variables gives the unconstrained optimal length
\begin{align}
    L^\star = \frac{1}{\chi_l}\left[ 2 + F^{-1}\left(\frac{\lambda_l\Lambda(f_l)}{2\beta_l\chi_l}\right)\right]. 
\end{align}
Finally, the beamwidth constraint requires $L \ge 0.886\frac{\lambda_l}{\theta_\epsilon}$. Incorporating this yields the complete optimal solution presented in \eqref{eq:optimalL}, which concludes the proof.

\subsection{Proof of \eqref{eq:BBR2}}
Similar to the proof of \eqref{eq:BBR1}, the derivation of BBR noise correlation function is presented in \eqref{ap:2}. Here, (a) arises because the BBR fields among different cells are uncorrelated.

\subsection{Proof of \eqref{eq:optimalM}}
Through algebraic manipulation, the objective function in \eqref{eq:P1} can be expressed in normalized form as
\begin{align}
    g(\bar{M}) = \frac{\bar{M}}{\bar{M}e^{\bar{M}}+b},
\end{align}
where $\bar{M}:= 2|\log n_t|M$ and $b:= \frac{2|\log n_t|L^2e^{-\chi_l L}\Lambda(f_l)}{\beta_l}$. To find the optimum, we examine the derivative of $g(\bar{M})$: 
\begin{align}\label{eq:grad_g}
    \frac{{\rm d} g(\bar{M})}{{\rm d}\bar{M}} = \frac{b - \bar{M}^2e^{\bar{M}}}{(\bar{M}e^{\bar{M}}+b)^2}.
\end{align}
Setting the numerator to zero yields the condition for a critical point: $b = \bar{M}^2e^{\bar{M}}$. Because the function $x^2e^x$ is strictly increasing for $x > 0$, $b = \bar{M}^2e^{\bar{M}}$ has a unique positive solution, written as 
\begin{align}
    \bar{M}^\star = S^{-1}(b),
\end{align}
where $S(x) := x^2e^x$. From \eqref{eq:grad_g}, we observe that $ \frac{{\rm d} g(\bar{M})}{{\rm d}\bar{M}} > 0$ when $\bar{M}<\bar{M}^\star$ and $ \frac{{\rm d} g(\bar{M})}{{\rm d}\bar{M}} < 0$ when $\bar{M}>\bar{M}^\star$.  
Hence, $g(\bar{M})$ is unimodal and achieves its global maximum at $\bar{M}^\star$
Reverting to the original variables gives the optimal number of segments in \eqref{eq:optimalM}, which completes the proof.


\bibliographystyle{IEEEtran}
\bibliography{Reference.bib}

\end{document}

%% file: header.tex
\newtheorem{theorem}{Theorem}
\newtheorem{acknowledgement}[theorem]{Acknowledgement}
\newtheorem{axiom}[theorem]{Axiom}
\newtheorem{case}[theorem]{Case}
\newtheorem{claim}[theorem]{Claim}
\newtheorem{conclusion}[theorem]{Conclusion}
\newtheorem{condition}[theorem]{Condition}
\newtheorem{conjecture}[theorem]{Conjecture}
\newtheorem{criterion}[theorem]{Criterion}
\newtheorem{definition}{Definition}
\newtheorem{exercise}[theorem]{Exercise}
\newtheorem{lemma}{Lemma}
\newtheorem{corollary}{Corollary}
\newtheorem{notation}[theorem]{Notation}
\newtheorem{problem}[theorem]{Problem}
\newtheorem{proposition}{Proposition}
\newtheorem{solution}[theorem]{Solution}
\newtheorem{summary}[theorem]{Summary}
\newtheorem{assumption}{Assumption}
\newtheorem{example}{\bf Example}
\newtheorem{remark}{\bf Remark}

\newtheorem{thm}{Corollary}[section]
\renewcommand{\thethm}{\arabic{section}.\arabic{thm}}

\def\qed{$\Box$}
\def\QED{\mbox{\phantom{m}}\nolinebreak\hfill$\,\Box$}
\def\proof{\noindent{\emph{Proof:} }}
\def\poof{\noindent{\emph{Sketch of Proof:} }}
\def
\endproof{\hspace*{\fill}~\qed
\par
\endtrivlist\unskip}
\def\endproof{\hspace*{\fill}~\qed\par\endtrivlist\vskip3pt}

\def\E{\mathsf{E}}
\def\eps{\varepsilon}
\def\phi{\varphi}
\def\Lsp{{\boldsymbol L}}
\def\Bsp{{\boldsymbol B}}
\def\lsp{{\boldsymbol\ell}}
\def\Ltsp{{\Lsp^2}}
\def\Lpsp{{\Lsp^p}}
\def\Linsp{{\Lsp^{\infty}}}
\def\LtR{{\Lsp^2(\Rst)}}
\def\ltZ{{\lsp^2(\Zst)}}
\def\ltsp{{\lsp^2}}
\def\ltZt{{\lsp^2(\Zst^{2})}}
\def\ninN{{n{\in}\Nst}}
\def\oh{{\frac{1}{2}}}
\def\grass{{\cal G}}
\def\ord{{\cal O}}
\def\dist{{d_G}}
\def\conj#1{{\overline#1}}
\def\ntoinf{{n \rightarrow \infty}}
\def\toinf{{\rightarrow \infty}}
\def\tozero{{\rightarrow 0}}
\def\trace{{\operatorname{trace}}}
\def\ord{{\cal O}}
\def\UU{{\cal U}}
\def\rank{{\operatorname{rank}}}
\def\acos{{\operatorname{acos}}}

\def\SINR{\mathsf{SINR}}
\def\SNR{\mathsf{SNR}}
\def\SIR{\mathsf{SIR}}
\def\tSIR{\widetilde{\mathsf{SIR}}}
\def\Ei{\mathsf{Ei}}
\def\l{\left}
\def\r{\right}
\def\lb{\left\{}
\def\rb{\right\}}

\setcounter{page}{1}

\newcommand{\eref}[1]{(\ref{#1})}
\newcommand{\fig}[1]{Fig.\ \ref{#1}}

\def\bydef{:=}
\def\ba{{\mathbf{a}}}
\def\bb{{\mathbf{b}}}
\def\bc{{\mathbf{c}}}
\def\bd{{\mathbf{d}}}
\def\bee{{\mathbf{e}}}
\def\bff{{\mathbf{f}}}
\def\bg{{\mathbf{g}}}
\def\bh{{\mathbf{h}}}
\def\bi{{\mathbf{i}}}
\def\bj{{\mathbf{j}}}
\def\bk{{\mathbf{k}}}
\def\bl{{\mathbf{l}}}
\def\bm{{\mathbf{m}}}
\def\bn{{\mathbf{n}}}
\def\bo{{\mathbf{o}}}
\def\bp{{\mathbf{p}}}
\def\bq{{\mathbf{q}}}
\def\br{{\mathbf{r}}}
\def\bs{{\mathbf{s}}}
\def\bt{{\mathbf{t}}}
\def\bu{{\mathbf{u}}}
\def\bv{{\mathbf{v}}}
\def\bw{{\mathbf{w}}}
\def\bx{{\mathbf{x}}}
\def\by{{\mathbf{y}}}
\def\bz{{\mathbf{z}}}
\def\b0{{\mathbf{0}}}

\def\bA{{\mathbf{A}}}
\def\bB{{\mathbf{B}}}
\def\bC{{\mathbf{C}}}
\def\bD{{\mathbf{D}}}
\def\bE{{\mathbf{E}}}
\def\bF{{\mathbf{F}}}
\def\bG{{\mathbf{G}}}
\def\bH{{\mathbf{H}}}
\def\bI{{\mathbf{I}}}
\def\bJ{{\mathbf{J}}}
\def\bK{{\mathbf{K}}}
\def\bL{{\mathbf{L}}}
\def\bM{{\mathbf{M}}}
\def\bN{{\mathbf{N}}}
\def\bO{{\mathbf{O}}}
\def\bP{{\mathbf{P}}}
\def\bQ{{\mathbf{Q}}}
\def\bR{{\mathbf{R}}}
\def\bS{{\mathbf{S}}}
\def\bT{{\mathbf{T}}}
\def\bU{{\mathbf{U}}}
\def\bV{{\mathbf{V}}}
\def\bW{{\mathbf{W}}}
\def\bX{{\mathbf{X}}}
\def\bY{{\mathbf{Y}}}
\def\bZ{{\mathbf{Z}}}

\def\bxi{{\boldsymbol{\xi}}}

\def\sT{{\mathsf{T}}}
\def\sH{{\mathsf{H}}}
\def\cmp{{\text{cmp}}}
\def\cmm{{\text{cmm}}}
\def\WPT{{\text{WPT}}}
\def\lo{{\text{lo}}}
\def\gl{{\text{gl}}}

\def\tT{{\widetilde{T}}}
\def\tF{{\widetilde{F}}}
\def\tP{{\widetilde{P}}}
\def\tG{{\widetilde{G}}}
\def\tbh{{\widetilde{\mathbf{h}}}}
\def\tbg{{\widetilde{\mathbf{g}}}}

\def\mA{{\mathbb{A}}}
\def\mB{{\mathbb{B}}}
\def\mC{{\mathbb{C}}}
\def\mD{{\mathbb{D}}}
\def\mE{{\mathbb{E}}}
\def\mF{{\mathbb{F}}}
\def\mG{{\mathbb{G}}}
\def\mH{{\mathbb{H}}}
\def\mI{{\mathbb{I}}}
\def\mJ{{\mathbb{J}}}
\def\mK{{\mathbb{K}}}
\def\mL{{\mathbb{L}}}
\def\mM{{\mathbb{M}}}
\def\mN{{\mathbb{N}}}
\def\mO{{\mathbb{O}}}
\def\mP{{\mathbb{P}}}
\def\mQ{{\mathbb{Q}}}
\def\mR{{\mathbb{R}}}
\def\mS{{\mathbb{S}}}
\def\mT{{\mathbb{T}}}
\def\mU{{\mathbb{U}}}
\def\mV{{\mathbb{V}}}
\def\mW{{\mathbb{W}}}
\def\mX{{\mathbb{X}}}
\def\mY{{\mathbb{Y}}}
\def\mZ{{\mathbb{Z}}}

\def\cA{\mathcal{A}}
\def\cB{\mathcal{B}}
\def\cC{\mathcal{C}}
\def\cD{\mathcal{D}}
\def\cE{\mathcal{E}}
\def\cF{\mathcal{F}}
\def\cG{\mathcal{G}}
\def\cH{\mathcal{H}}
\def\cI{\mathcal{I}}
\def\cJ{\mathcal{J}}
\def\cK{\mathcal{K}}
\def\cL{\mathcal{L}}
\def\cM{\mathcal{M}}
\def\cN{\mathcal{N}}
\def\cO{\mathcal{O}}
\def\cP{\mathcal{P}}
\def\cQ{\mathcal{Q}}
\def\cR{\mathcal{R}}
\def\cS{\mathcal{S}}
\def\cT{\mathcal{T}}
\def\cU{\mathcal{U}}
\def\cV{\mathcal{V}}
\def\cW{\mathcal{W}}
\def\cX{\mathcal{X}}
\def\cY{\mathcal{Y}}
\def\cZ{\mathcal{Z}}
\def\cd{\mathcal{d}}
\def\Mt{M_{t}}
\def\Mr{M_{r}}
\def\O{\Omega_{M_{t}}}
\newcommand{\figref}[1]{{Fig.}~\ref{#1}}
\newcommand{\tabref}[1]{{Table}~\ref{#1}}

\newcommand{\fb}{\tx{fb}}
\newcommand{\nf}{\tx{nf}}
\newcommand{\BC}{\tx{(bc)}}
\newcommand{\MAC}{\tx{(mac)}}
\newcommand{\Pout}{p_{\mathsf{out}}}
\newcommand{\nnn}{\nn\\}
\newcommand{\FB}{\tx{FB}}
\newcommand{\TX}{\tx{TX}}
\newcommand{\RX}{\tx{RX}}
\renewcommand{\mod}{\tx{mod}}
\newcommand{\m}[1]{\mathbf{#1}}
\newcommand{\td}[1]{\tilde{#1}}
\newcommand{\sbf}[1]{\scriptsize{\textbf{#1}}}
\newcommand{\stxt}[1]{\scriptsize{\textrm{#1}}}
\newcommand{\suml}[2]{\sum\limits_{#1}^{#2}}
\newcommand{\sumlk}{\sum\limits_{k=0}^{K-1}}
\newcommand{\eqhsp}{\hspace{10 pt}}
\newcommand{\tx}[1]{\texttt{#1}}
\newcommand{\Hz}{\ \tx{Hz}}
\newcommand{\sinc}{\tx{sinc}}
\newcommand{\diag}{\mathrm{diag}}
\newcommand{\MAI}{\tx{MAI}}
\newcommand{\ISI}{\tx{ISI}}
\newcommand{\IBI}{\tx{IBI}}
\newcommand{\CN}{\tx{CN}}
\newcommand{\CP}{\tx{CP}}
\newcommand{\ZP}{\tx{ZP}}
\newcommand{\ZF}{\tx{ZF}}
\newcommand{\SP}{\tx{SP}}
\newcommand{\MMSE}{\tx{MMSE}}
\newcommand{\MINF}{\tx{MINF}}
\newcommand{\RC}{\tx{MP}}
\newcommand{\MBER}{\tx{MBER}}
\newcommand{\MSNR}{\tx{MSNR}}
\newcommand{\MCAP}{\tx{MCAP}}
\newcommand{\vol}{\tx{vol}}
\newcommand{\ah}{\hat{g}}
\newcommand{\tg}{\tilde{g}}
\newcommand{\teta}{\tilde{\eta}}
\newcommand{\heta}{\hat{\eta}}
\newcommand{\uh}{\m{\hat{s}}}
\newcommand{\eh}{\m{\hat{\eta}}}
\newcommand{\hv}{\m{h}}
\newcommand{\hh}{\m{\hat{h}}}
\newcommand{\Po}{P_{\mathrm{out}}}
\newcommand{\Poh}{\hat{P}_{\mathrm{out}}}
\newcommand{\Ph}{\hat{\gamma}}
\newcommand{\mat}[1]{\begin{matrix}#1\end{matrix}}
\newcommand{\ud}{^{\dagger}}
\newcommand{\C}{\mathcal{C}}
\newcommand{\nn}{\nonumber}
\newcommand{\nInf}{U\rightarrow \infty}